\documentclass[fleqn,usenatbib,useAMS]{mnras}

\usepackage{graphicx}	% Including figure files
\usepackage{amsmath}	% Advanced maths commands
\usepackage{amssymb}	% Extra maths symbols
\usepackage{multicol}        % Multi-column entries in tables
\usepackage{bm}		% Bold maths symbols, including upright Greek
\usepackage{pdflscape}	% Landscape pages

\usepackage{hyperref}
\usepackage{cleveref}

\crefformat{figure}{Fig.~#2#1#3}
\crefformat{equation}{equation~(#2#1#3)}
\crefformat{table}{Tab.~#2#1#3}
\crefformat{appendix}{appendix~#2#1#3}
\crefmultiformat{figure}{Figs.~#2#1#3}{~and~#2#1#3}%
    {,~#2#1#3}{,~#2#1#3}
\crefrangeformat{figure}{Figs.~(#3#1#4--#5#2#6)}

\usepackage[T1]{fontenc}
\usepackage{ae,aecompl}

\usepackage{newtxtext,newtxmath}
\title[Joint constraints on the first galaxies]{Joint analysis constraints on the physics of the first galaxies with low frequency radio astronomy data}

\author[Harry T. J. Bevins et al.]{Harry T. J. Bevins$^{1, 2}$\thanks{htjb2@cam.ac.uk}, Stefan Heimersheim$^{3}$, Irene Abril-Cabezas$^{3}$, 
Anastasia Fialkov$^{2, 3}$, \newauthor
Eloy de Lera Acedo$^{1, 2}$, William Handley$^{1, 2}$, Saurabh Singh$^{4}$ and Rennan Barkana$^{5, 6}$ \\
$^{1}$ Astrophysics Group, Cavendish Laboratory, J. J. Thomson Avenue, Cambridge, CB3 0HE, UK \\
$^{2}$ Kavli Institute for Cosmology, Madingley Road, Cambridge CB3 0HA, UK \\
$^{3}$ Institute of Astronomy, University of Cambridge, Madingley Road, Cambridge CB3 0HA, UK \\
$^{4}$Raman Research Institute, C V Raman Avenue, Sadashivanagar, Bangalore 560080, India \\
$^{5}$School of Physics and Astronomy, Tel-Aviv University, Tel-Aviv, 69978, Israel \\
$^{6}$Institute for Advanced Study, 1 Einstein Drive, Princeton, New Jersey 08540, USA}

\date{Last updated 2020 June 10; in original form 2013 September 5}

\pubyear{2023}

\begin{document}
\label{firstpage}
\pagerange{\pageref{firstpage}--\pageref{lastpage}}
\maketitle

\begin{abstract}
The first billion years of cosmic history remains largely unobserved.
We demonstrate, using a novel machine learning technique, how combining upper limits on the spatial fluctuations in the 21-cm signal with observations of the sky-averaged 21-cm signal from neutral hydrogen can improve our understanding of this epoch.
By jointly analysing data from SARAS3 (redshift $z\approx15-25$) and limits from HERA ($z\approx8$ and $10$), we show that such a synergetic analysis provides tighter constraints on the astrophysics of galaxies 200 million years after the Big Bang than can be achieved with the individual data sets. Although our constraints are weak, this is the first time data from a sky-averaged 21-cm experiment and power spectrum experiment have been analysed together. In synergy, the two experiments leave only $64.9^{+0.3}_{-0.1}$\% of the explored broad theoretical parameter space to be consistent with the joint data set, in comparison to $92.3^{+0.3}_{-0.1}$\% for SARAS3 and $79.0^{+0.5}_{-0.2}$\% for HERA alone. We use the joint analysis to constrain star formation efficiency, minimum halo mass for star formation, X-ray luminosity of early emitters and the radio luminosity of early galaxies. The joint analysis disfavours at 68\% confidence a combination of galaxies with X-ray emission that is $\lesssim 33$ and radio emission that is $\gtrsim 32$ times as efficient as present day galaxies. We disfavour at $95\%$ confidence scenarios %with radio-luminous galaxies 
in which power spectra are $\geq126$~mK$^{2}$ at $z=25$ and the sky-averaged signals are $\leq-277$~mK.
\end{abstract}


%%%%%%%%%%%%%%%%%%%%%%%%%%%%%%%%%%%%%%%%%%%%%%%%%%

%%%%%%%%%%%%%%%%% BODY OF PAPER %%%%%%%%%%%%%%%%%%

\section{Introduction}

The period of cosmic history covering the birth of primordial stars, formation of the very first black holes and assembly of the earliest galaxies is largely unconstrained by observations. Covering the first billion years after the formation of the Cosmic Microwave Background~(CMB), the period is currently out of reach of the modern telescopes, but theoretical models suggest the first star formed between $z\sim 20-60$, i.e. around $35-200$ million years after the Big Bang \citep{Bond_pop3_1981, Bromm_pop3_2004, Klessen_pop3_2019}.

Observational prospects of this epoch are improving with the exciting new observations from the James Webb Space Telescope \citep[JWST, ][]{Windhorst_JWST_2006, Park_JWST_2020, Robertson_JWST_2022}, which is expected to be able to observe bright galaxies out to redshift $z\approx20$ \citep[e.g.][]{Windhorst_JWST_2006, Naidu_2022, JADES_highz_2023} and is much more sensitive than the previous generation of telescopes such as the Hubble Space Telescope.
The highest spectroscopically confirmed galaxy observed with JWST at the time of writing is at $z=13.20$ \citep{Robertson_2023}. This picture of the infant Universe will potentially be supplemented by proposed X-ray observatories such as ATHENA \citep{Athena}, AXIS \citep{axis} and LYNX \citep{Lynx} that will probe X-ray sources and the hot diffused gas in the infant Universe, improving our understanding of some of the first X-ray emitting objects. In addition, radio telescopes such as the Square Kilometre Array \citep[SKA, ][]{Mellema_SKA_2013, Koopmans_SKA_2015, Greig_SKA_2020} will provide information about the state of the Intergalactic Medium through tomographic observations of the 21-cm line \citep{Furlanetto_review_2006, Barkana_review_2016, Mesinger_review_2019} during the Cosmic Dawn and Epoch of Reionization~(EoR). In the coming decades, through a joined effort over a range of different wavelengths, researchers will reveal the first billion years of cosmic history between the formation of the CMB and the current epoch of the galaxies.

In 2018 the EDGES collaboration reported a tentative detection of the sky-averaged (or global) 21-cm signal \citep{EDGES} which led to an increased interest in the field. The reported absorption feature is much deeper than conventional theoretical models \citep{Mesinger_21cmFAST_2011, Mirocha_global_2017, Cohen_global_2017, Reis_lya_2021, Munoz_first_gals_2022} and implies a rapid period of efficient star formation followed by a delayed onset of strong heating of the neutral gas. Exotic processes such as cooling of the gas through interactions with charged cold dark matter \citep{BarkanaDM2018, Barkana2018, Slatyer2018, Berlin2018, Kovetz2018, Munoz2018, Fialkov2018, Liu2019, axion_dm_2021, Wimpless_DM_2022, Caputo_2023} and models with excess radio backgrounds above the CMB \citep{Ewall2018, Jana2018, Fialkov2019, Reis2020, Mittal_ERB_2022, Acharya_2022, Acharya_2023} have been proposed to explain the depth of the absorption feature. However, explaining the implied rapid onset of star formation, delayed and then rapid heating is much harder and a number of works have suggested that there are unaccounted for systematics in the data \citep{Hills2018, Singh2019, Bradley_EDGES_2019, Sims2020, maxsmooth}.

A number of other experimental efforts are underway to try to detect the high-redshift 21-cm signal. Upper limits on the 21-cm power spectrum, which measures fluctuations in the 21-cm field of time and space, have been reported by PAPER \citep{Jacobs_Paper_limits_2015}, MWA \citep{Trott_2020, Kolopanis_MWA_limits_2022}, LOFAR \citep{Patil_2017, Mertens_2020}, AARTFAAC \citep{AARTFAAC_2020}, HERA \citep{HERA_2022a, HERA_2022b, HERA_2022c}, LWA \citep{LWA_PS_2019} and LEDA \citep{LEDA_PS_2021}. Similarly, limits on the magnitude of the global 21-cm signal have been reported by several experiments including SARAS \citep{SARAS2_2017, SARAS2_2018, Bevins_saras2_2022, Bevins_saras3_2022}, EDGES High Band \citep{Monsalve_EDGES_HB_1_2017, Monsalve_EDGES_HB_2_2018, Monsalve_EDGES_HB_3_2019} and LEDA \citep{Bernardi_LEDA_2016}. The limits have been used to place constraints on the properties of the first galaxies \citep{Mondal_LOFAR_2022, HERA_2022b, HERA_2022c, SARAS2_2017, SARAS2_2018, Monsalve_EDGES_HB_1_2017, Monsalve_EDGES_HB_2_2018, Monsalve_EDGES_HB_3_2019, Bevins_saras2_2022, Bevins_saras3_2022} and are improving significantly over time. Future space based missions such as DAPPER and FARSIDE \citep{Burns_2021} as well as ongoing and upcoming experiments from Earth such as SARAS \citep{SARAS3}, MIST \citep{MIST}, REACH \citep{REACH}, LOFAR \citep{LOFAR_EoR_2018}, NenuFAR \citep{Zarka_nenuFar_2018}, HERA \citep{DeBoer_HERA_2017}, the SKA all hope to further improve our understanding of the early Universe through the 21-cm line.
 
In this paper, we take upper limits from 18 nights of observations with the HERA interferometer on the magnitude of the 21-cm power spectrum at redshifts $z \approx 8$ and $\approx 10$. HERA provides a window to the EoR when the neutral hydrogen gas was efficiently ionized by ultraviolet emission from the first massive galaxies \citep{HERA_2022b}, and the SARAS3 radiometer \citep{SARAS3} that probes the sky-averaged 21-cm emission at redshifts between $z\approx 15 - 25$. SARAS3 observes the Cosmic Dawn when the first small galaxies, containing the first stars and X-ray emitting objects, are expected to have formed. For the first time, we jointly analyse the upper limits from HERA and SARAS3 to improve our understanding about the properties of the Intergalactic Medium during the Cosmic Dawn and the properties of the first galaxies. While in the HERA analysis the authors explore a number of different models of the 21-cm signal, we focus here on the models that include an excess radio background above the CMB from high redshift radio galaxies, Lyman-$\alpha$ heating, CMB heating, multiple scattering of Lyman-$\alpha$ photons and X-ray heating \citep[see ][]{Reis2020, Reis_lya_2021}. Since these models were also used in the SARAS3 analysis, it is convenient to use them here to showcase the power of the joint analysis. We refer to the HERA constraints on this specific 21-cm model as `the HERA results' throughout the work but stress that the constraints are model dependent, and we make efforts to discuss constraints on derived parameters such as the radio background temperature and spin temperature allowing for comparisons across different models of the 21-cm signal where possible. 

We introduce a machine learning enhanced Bayesian workflow to efficiently perform the joint analysis. When considered together, we find that the results from HERA and SARAS3 provide a better leverage on theoretical scenarios across a wider redshift range than the constraints from each of the individual experiment.

In \cref{sec:method} we review the data and the specifics of the experiments incorporated in our analysis. This is followed by a discussion of our machine learning enhanced Bayesian workflow and about the synergies between the power spectrum and sky-averaged 21-cm experiments. We present the implications of our work for the astrophysical constraints in \cref{sec:results}. This is followed by conclusions in \cref{sec:conclusions}.

\section{Methodology}\label{sec:method}

\subsection{Data}

Observing from a lake in Southern India, the global 21-cm experiment SARAS3 took 15 hours of usable data over the frequency band $55 - 85$ MHz. After correcting for the impact of the environment, man-made radio signals and the efficiency of the antenna, a measurement of the average sky temperature was attained. The data is expected to include foregrounds from the Galaxy and other galaxies as well as the 21-cm signal. The data was used to constrain the properties of the first galaxies \citep{Bevins_saras3_2022} and the collaboration reported a null detection of the EDGES absorption feature with 95.3\% confidence \citep{SARAS3}.

A previous iteration of the SARAS experiment, SARAS2, observed at higher frequencies, $110 - 200$ MHz. The data contained systematic structures, thought to be from the environment, but was used to place constraints on the magnitude of the 21-cm signal and properties of early galaxies \citep{SARAS2_2017, SARAS2_2018, Bevins_saras3_2022}. In this work, we combine constraints from \cite{Bevins_saras3_2022} with constraints from SARAS3 and HERA.

The best upper limits observed by interferometric experiments to date have come from HERA interferometer \citep{HERA_2022b},  followed by MWA in the redshift range $z = 6.5 - 8.7$ \citep{Trott_2020} and LOFAR, which provide the tightest upper limits at redshifts $z \sim 9.1$ \citep{Patil_2017} and $z \sim 9.3 - 10.6$ \citep{Mertens_2020}. HERA is located in South Africa in the Karoo Desert \citep{DeBoer_HERA_2017} and in this work we use spherically averaged power spectra upper limits from Internal Data Release 2 \citep{HERA_2022a} which include observations over 18 nights with 39 antennas. The data covers the wavenumber range $k=0.128\,h\mathrm{Mpc}^{-1}$ to $0.960\,h\mathrm{Mpc}^{-1}$ and from the two bands focusing on redshifts $z=7.9$ and $10.3$. We leave the inclusion of constraints from the analysis of Internal Data Release 3 which was recently published \citep{HERA_2022c} to future work, noting that there is an overlap between the data sets. In \cref{sec:mwa_lofar}, we consider the implications of including the upper limits on the power spectrum from MWA and LOFAR in our joint analysis with SARAS3 and HERA.

\subsection{Machine learning enhanced Bayesian workflow}

In 21-cm cosmology, data analysis efforts are increasingly employing Bayes theorem 
\begin{equation}
    \mathcal{P}(\theta|D, \mathcal{M}) = \frac{\mathcal{L}(\theta) \pi(\theta)}{\mathcal{Z}},
    \label{eq:bayes}
\end{equation}
to derive constraints on the astrophysical scenarios of the early Universe. Here the likelihood, $\mathcal{L}(\theta)$, represents the probability that we observe the data, $D$, from SARAS3 or HERA, given a particular model, $\mathcal{M}$. The prior, $\pi(\theta)$, represents our assumed knowledge before we consider any data and the evidence, $\mathcal{Z}$, is a normalisation constant that can be used for model comparisons. 
The posterior, $\mathcal{P}(\theta|D, \mathcal{M})$, tells us which parts of the parameter space, $\theta$, given the data and chosen model, are more probable than others. %The evaluation of Bayes theorem is usually performed with Nested Sampling or Markov Chain Monte Carlo~(MCMC) algorithms (see \textit{Supplementary material}). 
In many 21-cm experiments, $\theta$ is composed of parameters that describe instrumental effects, $\theta_{I}$, foregrounds, $\theta_\mathrm{fg}$, and the astrophysical processes that influence the 21-cm signal, $\theta_{21}$. We typically refer to the set of $\theta_{I}$ and $\theta_\mathrm{fg}$ as the nuisance parameters. Since we are only interested in the 21-cm signal, we work with the marginal or nuisance-free posteriors and nuisance-free likelihoods, $\mathcal{L}(\theta_{21})$, which can be estimated using normalising flows and the marginal Bayesian statistics code \textsc{margarine} \citep{margarine, margarine2}. %We give an overview of this in \textit{Supplementary material}, however we note that it allows for efficient combination of the HERA and SARAS3 constraints.

%We use the marginal Bayesian statistics code \textsc{margarine} \citep{margarine} to combine constraints from different data sets. 
\textsc{margarine} uses neural networks known as Masked Autoregressive Flows (MAFs) to model the probability distribution, $\mathcal{P}(\theta|D, \mathcal{M})$, of a set of samples, here~{$\theta_{21}$}, marginalising over nuisance parameters describing the instrumental systematics, $\theta_{I}$, and foregrounds, $\theta_\mathrm{fg}$ in the process. It does this by shifting and scaling a base standard normal distribution, making the probability of the distribution easily tractable, to replicate the target samples, where the shifting and scaling are determined by the outputs of the MAFs. 

This can subsequently be used to calculate a nuisance-free likelihood given a prior distribution and the Bayesian evidence using
\begin{equation}
    \mathcal{L}(\theta_{21}) 
\equiv \frac{\int\mathcal{L}(\theta_{21},\alpha)\pi(\theta_{21},\alpha)d\alpha}{\int \pi(\theta_{21},\alpha)d\alpha} = \frac{\mathcal{P}(\theta_{21}|D, \mathcal{M})\mathcal{Z}}{\pi(\theta_{21})},
    \label{eqn:partial}
\end{equation}
where $\alpha = \{\theta_{I}, \theta_\mathrm{fg}\}$ \citep{margarine2}. In instances when the prior is flat then the following is true up to a normalisation constant
\begin{equation}
    \mathcal{L}(\theta_{21}) \approx \mathcal{P}(\theta_{21}|D, \mathcal{M}).
\end{equation}

With \textsc{margarine} we can then evaluate $\log(\mathcal{L}(\theta_{21}))$ for any set of $\theta_{21}$ for any existing posterior distribution, $\mathcal{P}(\theta|D, \mathcal{M})$, from previous analysis of a data set like HERA or SARAS3
\begin{equation}
\begin{split}
    \theta = \{\theta_{I}, \theta_\mathrm{fg}, \theta_{21}\} \rightarrow &\{\theta_{21}\}
    \rightarrow \textsc{margarine} \\
    & \rightarrow \log \mathcal{P}(\theta_{21}|D, \mathcal{M}) \rightarrow \log \mathcal{L}(\theta_{21}),
\end{split}
\end{equation}
where the $\log$ is base 10. The likelihood evaluations can then be combined,
\begin{equation}
    \log \mathcal{L}_{\textnormal{joint}}(\theta_{21}) = \log \mathcal{L}_{\textnormal{HERA}}(\theta_{21}) + \log \mathcal{L}_{\textnormal{SARAS3}}(\theta_{21}),
\end{equation}
to be sampled over using MCMC methods or in our case Nested Sampling implemented with \textsc{polychord} \citep{Handley2015a, Handley2015b}. 

MCMC sampling methods approximate the unnormalised posterior distribution by directly sampling the product $\mathcal{P}(\theta) \approx \mathcal{L}(\theta)\pi(\theta)$ and do not provide estimates of the evidence, $\mathcal{Z}$. The family of algorithms typically use random walkers to traverse the parameter space, with points accepted and rejected based on some probabilistic criteria in an effort to explore the space fully. The HERA analysis \citep{HERA_2022b} of the excess radio background models explored here used the \textsc{emcee} \citep{Foreman_Mackey_2013} implementation of MCMC sampling \citep{HERA_2022b}.

Nested sampling \citep{skilling_nested_2004} numerically approximates the integral
\begin{equation}
    \mathcal{Z} = \int \mathcal{L}(\theta)\pi(\theta) \delta \theta,
\end{equation}
which can be derived from \cref{eq:bayes} and the requirement that the posterior must integrate to 1, by evolving a series of live points to higher and higher likelihood values. Through approximating the evidence, the algorithm produces samples on the normalised posterior distribution, which we subsequently use to determine preferred and disfavoured regions of the parameter space. A comprehensive review of Nested Sampling can be found in \citep{Ashton_ns_review_2022}.

\subsection{Signal Modelling}

In order to realistically model the range of time covered by the Cosmic Dawn and Epoch of Reionization, we need a consistent modelling of the cosmological and astrophysical processes from redshift 60, when star formation might have began, all the way to redshift 5 at the end of the EoR. To that end we employ semi-numerical simulations \citep{Visbal_2012,Fialkov_lyw_2013, Fialkov_2014, Fialkov_rich_2014, Reis_lya_2021} that have previously been used in the HERA \citep{HERA_2022b}, SARAS2 \citep{Bevins_saras2_2022} and SARAS3 \citep{Bevins_saras3_2022} analyses as well as similar analysis of the LOFAR \citep{Mondal_LOFAR_2022} and EDGES High-Band \citep{Monsalve_EDGES_HB_3_2019} data. We model the three-dimensional 21-cm field as a function of cosmic time during the infant Universe taking into account important astrophysical process including the Wouthuysen-Field~(WF) effect  \citep{Wouthuysen, Field,  Fialkov_rich_2014}, heating of the intergalactic medium by X-ray \citep{Fialkov_2014}, Ly-$\alpha$ \citep{Reis_lya_2021} and CMB photons \citep{Fialkov2019}, multiple scattering of Ly-$\alpha$ \citep{Reis_lya_2021}, relative velocity between dark matter and gas \citep{Visbal_2012}, feedback of Lyman-Werner radiation on star formation \citep{Fialkov_lyw_2013},  and radio emission from galaxies \citep{Reis2020}.

At high redshifts around $z\approx 20 - 30$, the first stars begin to form and produce Ly-$\alpha$ photons that interact with the baryonic matter, predominantly composed of neutral hydrogen, in the Universe. Neutral hydrogen atoms absorb and re-emit ambient Ly-$\alpha$ photons in a process known as the Wouthuysen-Field~(WF) effect \citep{Wouthuysen, Field}, introduced above, that drives the relative number of atoms with aligned and anti-aligned proton and electron spins. This process couples the spin temperature of the neutral hydrogen, $T_s$, which describes the distribution of hydrogen atom spins, to the gas temperature, $T_k$, which is cooling at a faster rate than the radio background as the Universe expands. Further, interactions between the neutral hydrogen and Ly-$\alpha$ emission result in the transfer of kinetic energy that raises the gas temperature, and coupled spin temperature, in a process known as Ly-$\alpha$ heating \citep{Madau, Chuzhoy2007} preventing the gas from cooling adiabatically. However, despite the heating, the gas temperature remains cooler than the radio background at high redshifts, leading to an absorption feature in the sky-averaged 21-cm signal. Further, it leads to a peak in the power spectrum at high $z \approx 25$ on angular scales corresponding to the effective horizon of the Ly-$\alpha$ emission and the distribution of galaxies, which disappears when the coupling becomes saturated \citep{Cohen_power_2018}. The intensity and spatial fluctuations of the Ly-$\alpha$ emission evolve with the population of early galaxies and consequently it is dependent on their star formation rate and the minimum halo mass for star formation. We parameterize these quantities with the star formation efficiency, $f_*$, which quantifies the percentage of the baryonic mass in the star forming halos that is converted into stars, and the minimum circular velocity, $V_c$,  which is proportional to the cube root of the minimum halo mass for star formation, $M$.

At intermediate redshifts of $z\approx 10 -20$ the gas is further heated by X-ray binaries \citep{Fragos_Xrays_2013, Fialkov_2014}, continuing Ly-$\alpha$ heating, CMB heating \citep{Venumadhav2018} and heating through structure formation. X-ray heating is dependent on processes such as black hole binary formation and X-ray production in high redshift galaxies. This means that it has to be separately parameterized and in our semi-numerical simulations, we model the X-ray production efficiency, $f_X$, which is directly proportional to the X-ray luminosity per star formation rate, $L_X/$SFR measured in erg~s$^{-1}$~M$_\odot^{-1}$~yr, between 0.2 and 95~keV. The heating affects the redshift of the minimum and depth of the sky-averaged 21-cm signal. If sufficiently efficient it can raise the temperature of the gas, and, consequently, the coupled 21-cm brightness temperature, above the radio background resulting in an emission feature in the sky-averaged signal at low redshifts. The heating rate has a direct impact on how fast the brightness temperature transforms from absorption to 0~K or emission. In the power spectrum, due to the non-uniformity of heating, the various mechanisms can produce a peak in the signal at around $z\approx 15$. Although this redshift is model dependent \citep[see][]{Cohen_power_2018} and in some cases when heating is done by hard X-rays (energies $>1$ keV with long mean free paths) X-ray heating is smooth and no peak is imprinted in the power spectrum \citep{Fialkov_2014, Fialkov_rich_2014}.

Finally, at more recent times, $z \approx 5 - 15$, ultraviolet emission from the first massive galaxies begins to ionize the neutral hydrogen, stripping the abundant gas of its electrons. This reduces the sky-averaged 21-cm signal and when reionization is complete, the signal disappears. The process produces a peak in the power spectrum at the scales corresponding to the typical size of ionized bubbles, but again the signal is destroyed once reionization is complete. The process of reionization is highly dependent on the ionizing efficiency of sources, $\zeta$, which in the models explored here is normalized by the CMB optical depth, $\tau$. While $\tau$ has been weakly constrained by cosmological experiments such as Planck \citep{Planck2018}, we treat it as a free parameter and note that 21-cm cosmology offers a means by which to break degeneracies between $\tau$ and other cosmological parameters.

Throughout our analysis, we explore a broad range of radio luminosities that produce radio excesses above the CMB of between $\approx0.5 - 270$ times the CMB temperature at $z=20$ and $\approx1 - 32000$ times the CMB at $z=10$. While moderate radio emission might be expected \citep{Mirocha2019}, the extreme values of radio production efficiencies are not expected in nature and are usually invoked to explain the anomalously deep EDGES signal \citep{EDGES, Feng2018, Jana2018, Ewall2018, Mirocha2019, Fialkov2019, Reis2020} but struggle to explain the rapid star formation and rapid heating of the gas \citep{Mittal2022} that is implied by the shape of the EDGES signal.

To summarise, the key parameters in the astrophysical model are: the star formation efficiency, $f_*$, which quantifies the percentage of the baryonic mass in the star forming halos that is converted into stars;  the minimum virial circular velocity, $V_c$, which is proportional to the cube root of the halo mass, $M$; the X-ray production efficiency, $f_X$, which is directly proportional to the X-ray luminosity per star formation rate, $L_X/$SFR, measured in erg~s$^{-1}$~M$_\odot^{-1}$~yr, between 0.2 and 95~keV;  the CMB optical depth, $\tau$; finally, the radio production efficiency, $f_\mathrm{radio}$, of high redshift radio galaxies which is proportional to the radio luminosity per star formation rate, $L_r$, measured in W~Hz$^{-1}$~M$_\odot^{-1}$~yr at 150~MHz,
\begin{equation}
    L_\mathrm{r} = f_\mathrm{radio} 10^{22} \bigg(\frac{\nu}{150~\mathrm{MHz}}\bigg)^{-\alpha_\mathrm{radio}} \frac{\mathrm{SFR}}{\mathrm{M}_\odot\mathrm{yr}^{-1}},
    \label{eq:radio_luminosity}
\end{equation}
and normalized such that it has a value of one for the present day population of radio galaxies \citep{Reis2020}. In our Bayesian analysis, $\theta_{21}$, therefore, comprises the set of parameters \{$f_*, V_c, f_X, \tau, f_\mathrm{radio}$\} or equivalently \{$f_*, M, L_X/\textnormal{SFR}, \tau, L_\mathrm{r}/\textnormal{SFR}$\}.

We explore wide prior ranges on all the parameters in an attempt to let the data inform us about the high-redshift astrophysical processes. Specifically, we constrain $f_*$ in the range $0.001-0.5$, $V_c$ in the range $4.2 - 100$, $f_X$ in the range $10^{-4} - 1000$, $\tau$ in the range $0.035 - 0.077$ and finally $f_\mathrm{radio}$ in the range $1 - 99750$. Typically, the radio background is assumed to be equal to the CMB and contributions from radio-luminous galaxies are not conventionally considered. Here, we expect that early galaxies will contribute to the radio background, thus  increasing the amplitude of both the sky-averaged 21-cm signal \citep{Feng2018} and the power spectrum \citep{Fialkov2019, Reis2020}. 

Both the sky-averaged 21-cm signal and the power spectrum rely on the same underlying physics, and constraints from experiments targeting the different probes can be effectively combined to improve our knowledge of the infant Universe.

\subsection{Emulators}

The semi-numerical simulations used to model the signal take of order a few hours per parameter set, which is impractical for Nested Sampling or MCMC runs which need of order millions of signal evaluations (the SARAS3 constraints used in this work required $\approx 15$ million signal evaluations to derive). However, as both the sky-averaged signal and the power spectrum change smoothly when we vary parameters, we can interpolate values at intermediate parameters from existing simulations. An increasingly common practice to achieve this are the fast and precise neural network based emulators. Typically, these networks take of order a few tens of milliseconds to evaluate meaning they are much more well suited for computationally intensive fitting algorithms. The specific emulators used in this analysis are detailed below. 

In practice, it is important to include a theory error associated with the accuracy of our emulators in the likelihood functions and the uncertainty in the theoretical modelling \citep{greig2017simultaneously}. An emulator error was included in the HERA likelihood in the original analysis \citep{HERA_2022b} and is done when analysing data from MWA and LOFAR in \cref{sec:mwa_lofar} but not done in the original SARAS2 and SARAS3 analysis. Since we are using the posterior distributions from the original HERA, SARAS2 and SARAS3 analysis and density estimators to produce our marginal or nuisance-free likelihood functions, we are unable to include the emulator error in the SARAS2 and 3 likelihoods here. However, the error in the global signal emulator is lower than the noise in the SARAS3 and SARAS2 data and the noise is expected to dominate over the both the uncertainty in the theoretical modelling and the emulator error. Theory errors should be included in future analysis, particularly as limits on the magnitude of the power spectrum and global signal improve, before performing joint analysis.

\subsubsection{The sky-averaged 21-cm signal}

To emulate the sky-averaged 21-cm signal, we use the publicly available emulator \textsc{globalemu} \citep{globalemu} trained on sets of astrophysical simulations. %The simulations include the effects of Ly-$\alpha$ and CMB heating. The mean free path of ionizing photons, $R_\mathrm{mfp}$, is fixed at $40$~Mpc and the and X-ray heating is powered by a population of X-ray binaries with a realistic spectral energy distribution \citep{Fialkov_2014}. The models correspond to the parameterisation detailed above and in \citep{Reis2020}.

The emulator has previously been used in the individual analysis of data from SARAS2 \citep{Bevins_saras2_2022} and SARAS3 \citep{Bevins_saras3_2022} and detailed in the corresponding papers. We therefore only briefly summarise the accuracy of the neural network here. The original set of simulations comprise approximately 10,000 models however the explored range of $\tau$, the CMB optical depth, is large and when training our neural network we filter out models that have a value of $\tau$ in the range given by Planck $\pm 3\sigma$. This results in a training set of approximately 4,300 models and the network is tested on approximately 500 models. For the test data set, a root mean squared error~(RMSE) of 5.11~mK is found and a 95 percentile RMSE of 20.53~mK indicating a high level of accuracy~\citep[see Table M.2 in ][]{Bevins_saras3_2022}.

\subsubsection{The 21-cm power spectrum}

For comparing models with the HERA measurements we use the same 21-cm power spectrum emulator used in the HERA analysis \citep{HERA_2022b}, trained on the same suit of simulations that give rise to the sky-averaged signal used in the SARAS2 and SARAS3 analysis.
Based on an input of the five model parameters, this emulator returns the 21-cm power spectrum with a relative accuracy of 20\% at the wave numbers and redshifts observed by HERA. %To cover the large dynamic range the emulator predicts
%$\log{\left(\Delta_{21}^2 + 1\,\mathrm{mK}^2\right)}$, as our instruments cannot detect signals $<1\,\mathrm{mK}^2$.
For HERA the impact of $\tau$ is more significant because the data corresponds to lower redshifts and so the full prior range on the parameter is used for training. This results in a training set of $\sim$8,000 simulations and another 2,000 independent samples for testing. Full details and tests of this emulator can be found in the HERA analysis paper \citep{HERA_2022b}, Appendix B.

We note that when performing our joint analysis we use the narrow prior on $\tau$ defined by Planck, since the SARAS3 posterior is not defined in the original analysis for values outside this range.

\subsubsection{Temperatures}

To emulate physical properties such as the spin temperature and temperature of the radio background (as seen in Figure \ref{fig:igm_params}) we use a \textsc{globalemu}-style emulator. In this framework, the emulator takes in the five astrophysical parameters and a single redshift and returns a single corresponding temperature. In practice, this means that vectorised calls have to be made to emulate the spin temperature, $T_s(z)$, and the radio background temperature, $T_r(z)$, as a function of redshift, but the method is found to be more accurate, quicker and allows for interpolation at a range of different redshifts. We use the same training and test sets as used for the 21-cm power spectrum emulator, and a similar architecture of 4 layers, with a reduced size of 100, 30, 10, and 5 nodes per layer. This allows us to emulate the spin temperature $T_s$ within $\pm 6\%$ accuracy (95\% confidence interval), and the radio background temperature $T_{r}$ within $\pm 4\%$ accuracy (95\% confidence interval).

\section{Results}\label{sec:results}

Although the constraints presented here are weak and model dependent, through the novelty of combining the previously reported HERA and SARAS3 constraints we produce the tightest constraints to date on the properties of the infant Universe, as detailed below. This is the first time a joint analysis between a global signal data and interferometric limits has been attempted. 

\subsection{The ratio of $T_s$ and $T_r$}

\begin{figure*}
    \centering
    \includegraphics{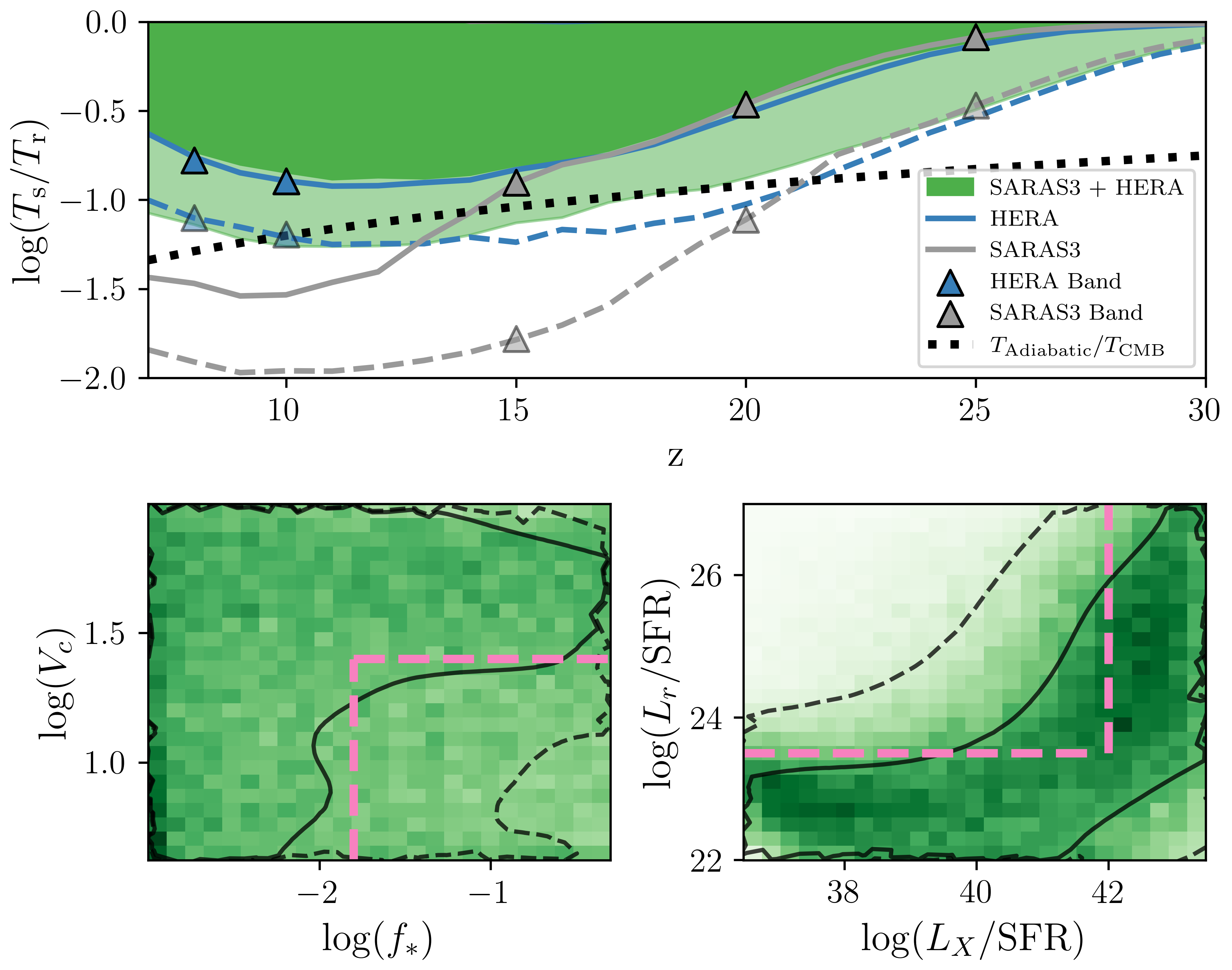}
    \caption{\textbf{Key results from the joint analysis.} \textbf{Top Panel:} The ratio of the spin temperature of neutral hydrogen, $T_s$, and the radio background temperature, $T_r$, as a function of redshift for the joint HERA and SARAS3 analysis in green. We show the HERA and SARAS3 68\% and 95\% confidence constraints in blue and grey respectively as triangles at the relevant redshifts and solid and dashed lines. %Typically, contours like those presented here are extrapolated outside observations, however, here the joint analysis results are interpolated between the lowest HERA redshift, $z=8$, and highest SARAS3 redshift, $z=25$. 
    As a guideline, we show the ratio for $T_r = T_\mathrm{CMB}$ and assuming  adiabatically cooled gas in an expanding Universe in the absence of any heating but with saturated coupling between $T_s$ and the gas kinetic temperature (dashed black line). 
    \textbf{Bottom Left:} The 2D PDF from the joint analysis on the minimum virial circular velocity, $V_c$, in combination with the star formation efficiency, $f_*$, marginalising over $f_\mathrm{radio}$, $f_X$ and $\tau$. The solid black line shows the 68\% contour, approximated by the pink dashed line, and the black dashed line shows the 95\% contour. The joint analysis disfavours low values of $V_c$ and high $f_*$ corresponding to efficient star formation. \textbf{Bottom right:} The constraint on the X-ray and radio luminosities from the joint analysis marginalising over $V_c$, $f_*$ and $\tau$. The joint analysis disfavours at 68\% confidence low X-ray efficiencies in combination with high radio production efficiencies.}
    \label{fig:igm_params}
\end{figure*}

To visualize the importance of combining the constraining power of HERA and SARAS3 we show, in the top panel of \cref{fig:igm_params}, constraints on the ratio of the spin temperature of neutral hydrogen, $T_s$, and background radiation temperature, $T_r$. The background radiation temperature is the sum of the CMB temperature and radio background from galaxies. The ratio determines the maximum absorption of the sky-averaged 21-cm signal, the smaller the ratio the larger the signal can be. The SARAS3 limits on the 21-cm signal (grey markers) correspond to lower limits on $T_s/T_r$, with the corresponding 1 and 2 $\sigma$ contours shown as lines and extrapolated out of the SARAS3 band. Our simulations provide a natural link between the power spectra and the global quantities, e.g. $T_s/T_r$, meaning that we can use the limits on the power spectrum from HERA to derive an equivalent constraint on $T_s/T_r$. These constraints are shown in \cref{fig:igm_params} (blue markers and lines). The joint constraint, as shown by the green shaded regions, provides the strongest constraints on the ratio, and in particular at redshifts $z=15 - 20$, gives better constraints than either experiment alone. Further, the combination of the two experimental data sets improves the constraints at intermediate redshifts over a pure extrapolation of each one of the sets of constraints. As a guideline, the dashed black line in the figure shows the ratio for $T_r = T_{CMB}$, i.e. in the absence of radio emission from galaxies, and assuming  adiabatically cooled gas  in an expanding Universe in the absence of any astrophysical heating sources but with saturated coupling between the 21-cm spin temperature and the gas kinetic temperature. This limit is often used in the literature to give context to the constraints \citep[e.g.][]{HERA_2022b, HERA_2022c}. However, we note that for the models tested here the gas does not cool adiabatically because of the CMB and Ly-$\alpha$ heating at early times, and we have an excess radio background above the CMB.

\subsection{Functional constraints on $\Delta^2$ and $T_{21}$}

\begin{figure*}
    \centering
    \includegraphics{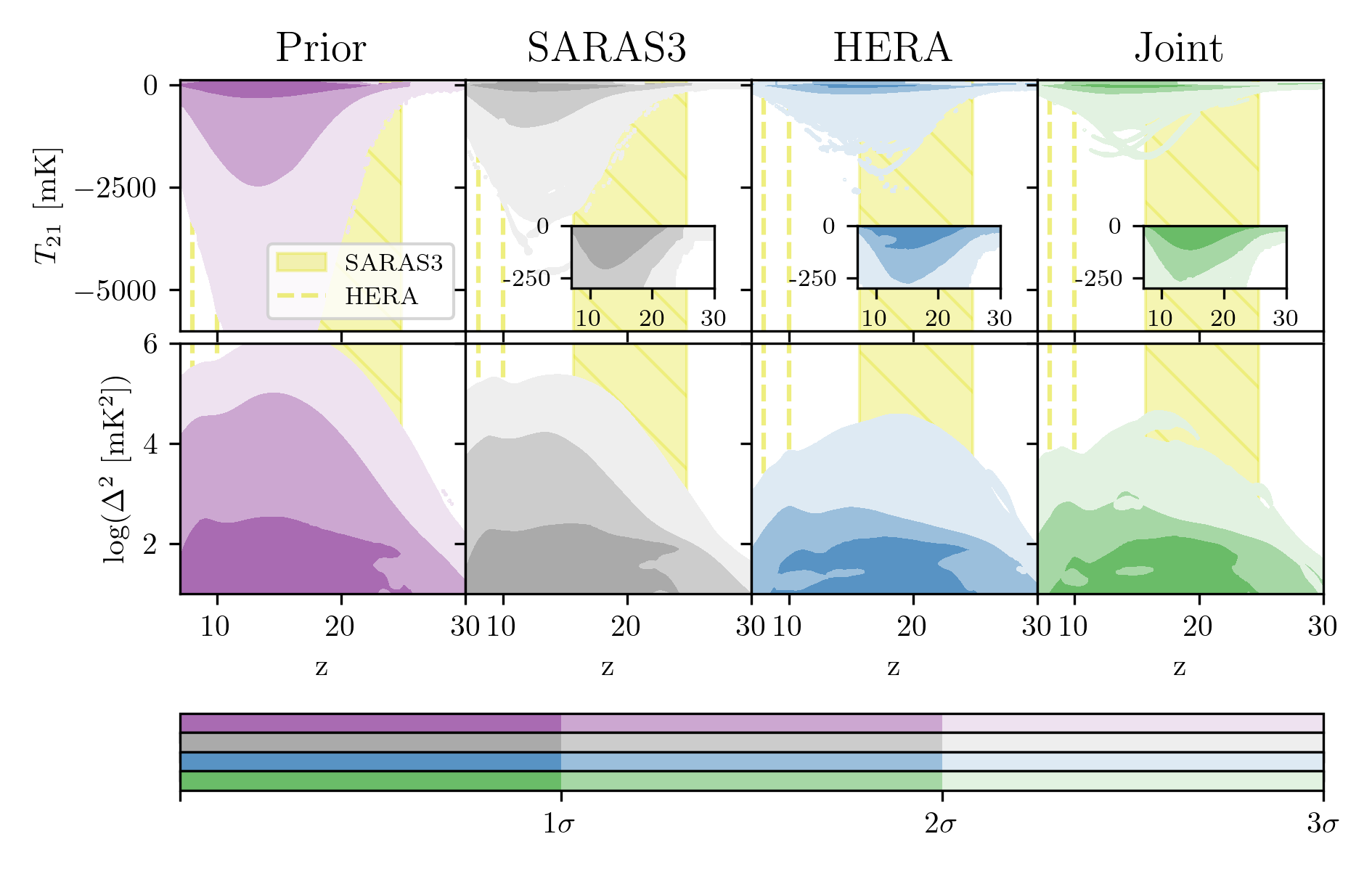}
    \caption{\textbf{Functional constraints on $T_{21}$ and $\Delta^2$.} The functional prior (purple), SARAS3 (grey), HERA (blue) and joint (green) posteriors for the sky-averaged 21-cm signal (top row) and power spectrum (bottom row). The yellow shaded region shows the SARAS3 band and the dashed yellow lines show the HERA redshifts. The functional prior and posteriors are calculated by taking representative samples from the corresponding probability distributions for the astrophysical parameters and generating the corresponding signals using neural networks. We see that by combining the constraining power of HERA and SARAS3, we significantly reduce the $3\sigma$ (lightest shaded regions) constraints on the magnitude of both signals from $-2630$~mK to $-1770$~mK at $z=15$ for the global signal and $10^{3.7}$~mK$^{2}$ to $10^{3.2}$~mK$^{2}$ for the power spectrum at $z=25$. The figure is produced with \textsc{fgivenx} \protect\citep{fgivenx}.}
    \label{fig:functional}
\end{figure*}

By taking the samples of $\theta_{21}$ output from our fits and using the neural network emulators to produce corresponding global signals and power spectra we can look at constraints on the temperature of the 21-cm signal and on the power spectrum. We explore the functional constraints in the $T_{21} - z$ and $\Delta^2 - z$ planes as shown in \cref{fig:functional}. We see that for both the power spectra and the sky-averaged signals, although more clearly for the latter, the range of plausible models is reduced by our joint analysis. The signals that are inconsistent with the data typically have strong power spectra and a corresponding deep absorption trough in the global signal, and belong to scenarios with weak X-ray heating and strong radio luminosity. Further, we see that the $2\sigma$ region for the functional constraints corresponds to signals with the power spectra  $\lesssim 10^{2.1}$ mK$^2$ at $z=25$, which via our modelling maps to global 21-cm signals shallower than $-277$~mK. We note that the $3\sigma$ limit on the power spectrum at $z=25$ of $\lesssim10^{3.2}$ mK$^2$ is approximately equivalent to the expected sensitivity of NenuFAR from 1000 hours of observations \citep{Mertens_NenuFAR_2021}. For the global 21-cm signal $3\sigma$ limit on the magnitude reduces from $\approx-2630$~mK from HERA to $\approx-1770$~mK from the joint analysis. Remarkably, we find that for the sky-averaged signal, the $2\sigma$ limit is very close to the minimum depth of the `standard astrophysical' models, $\approx -165$ mK \citep{Reis_lya_2021}, where the radio background is equated to the CMB, the contributions from radio galaxies and X-ray heating sources are assumed to be negligible, while the CMB and Ly-$\alpha$ heating are present. It is clear from our analysis, that the joint constraint improves limits on the range of plausible global  signals and  power spectra.

\subsection{Parameter Constraints}

\begin{figure*}
    \centering
    \includegraphics{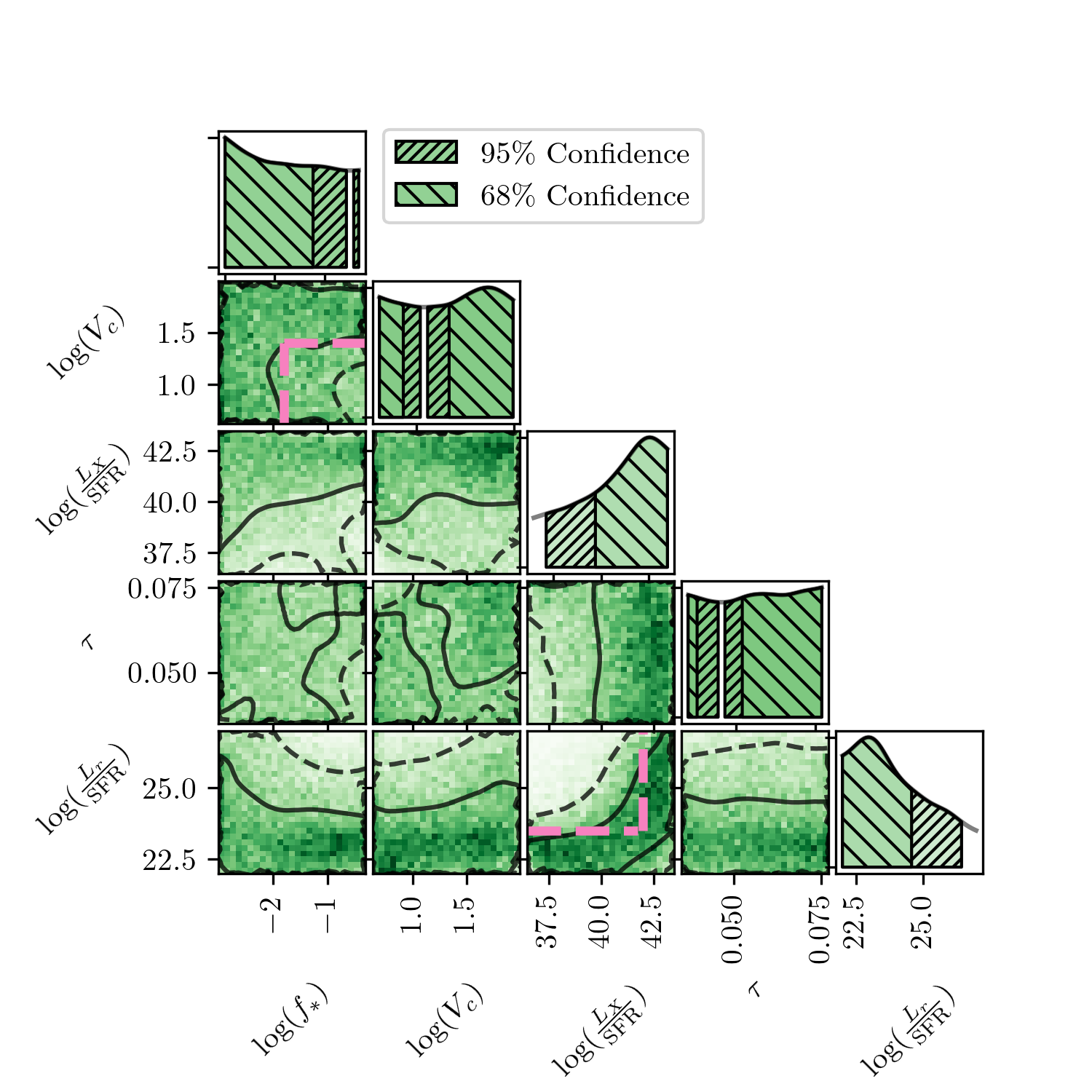}
    \caption{\textbf{Parameter constraints from the joint analysis.} The astrophysical parameter constraints on models with excess radio background above the CMB derived when combining an upper limit on the 21-cm power spectrum at $z\approx8$ and $\approx 10$ from HERA with data from the 21-cm sky-averaged experiment SARAS3 in the band $z\approx15-25$. Through the combination of these two data sets probing different statistical properties of the 21-cm signal at different redshifts we are able to improve constraints on the radio and X-ray luminosities of early radio-luminous galaxies and maintain constraints provided by SARAS3 on the star formation properties of these early galaxies. $L_r$ is measured in units of W~Hz$^{-1}$~M$_\odot^{-1}$~yr at 150~MHz and $L_X$ is in units of erg~s$^{-1}$~M$_\odot^{-1}$~yr calculated between 0.2 and 95~keV assuming a realistic SED of an early X-ray binary population. $V_c$ is measured in km/s. The pink dashed lines approximate regions that are disfavoured with 68\% confidence.}
    \label{fig:posterior}
\end{figure*}

We now interpret the constraints on the temperatures in terms of constraints on the properties of the first galaxies. 
The key constraints from the joint analysis are shown in the  bottom panels of \cref{fig:igm_params} including the limits  on the high-redshift star formation which drives the high-redshift portion of the 21-cm signal via the process of Ly-$\alpha$ coupling  (constraints in the plane $V_c - f_*$) and the constraints on the luminosity of X-ray and radio sources ($L_r - L_X$) which primarily regulates the depth of the absorption trough. The full marginalised 1D and 2D posteriors corresponding to the joint analysis are shown in \cref{fig:posterior} and the key numerical results are summarized in \cref{tab:numbers} alongside the individual constraints from SARAS3 and HERA. In \cref{app:key-params}, we show the individual constraints from each experiment and the joint constraint on $L_r - L_X$ and $V_c - f_*$ for comparison. We find that the combination of the two experiments leads to stronger constraints in the two-dimensional probability distribution of $L_r - L_X$ than either of the two experiments individually. Where HERA constrains the population of high redshift radio-luminous galaxies to be $\lesssim400$ times brighter in the radio band than the current population, the combination of the data sets constrains the galaxies to be $\lesssim300$ times brighter when marginalising over the other parameters ($V_c$, $f_*$, $f_X$ and $\tau$). Similarly, HERA disfavours at 68\% confidence galaxies with an X-ray luminosity $\lesssim 0.25$ times the present day value in combination with the radio luminosity of galaxies in the early universe that is $\gtrsim 400$ times the present day value. The joint analysis provides a stronger constraint, ruling out scenarios where the X-ray luminosity is $\lesssim 33$ times the present day value and the radio luminosity of the first galaxies is $\gtrsim32$ times the present day value at 68\% confidence.

\def\arraystretch{1.5}
\begin{table*}
    \centering
    \begin{tabular}{|p{2cm}|p{3.5cm}|p{3.5cm}|p{3.5cm}|}
         \hline
         & SARAS3 & HERA  & SARAS3 + HERA \\
         \hline
         \hline
         Signal & Sky-averaged & Power Spectrum & Both \\
         \hline
         $z$ & $\approx 15 - 25$ & $\approx 8$ \& $\approx 10$ & $\approx 8$, $\approx 10$ \& \newline $\approx 15 - 25$ \\
         \hline
         \hline
         $L_{r}/\mathrm{SFR}$ & $\gtrsim 1.55\times10^{25}$ & $\gtrsim4.00\times10^{24}$ & $\gtrsim3.31\times10^{24}$ \\
         \hline
         $L_{X}/\mathrm{SFR}$ & -- & $\lesssim7.60\times10^{39}$ & $\lesssim3.71\times10^{39}$ \\
         \hline
         $L_{r}/\mathrm{SFR}$ \& $L_{X}/\mathrm{SFR}$ & $\gtrsim 1.00\times10^{25}$ \& \newline $\lesssim 1.09\times10^{42}$ & $\gtrsim 4.00\times10^{24}$ \& \newline $\lesssim 7.60\times10^{39}$ & $\gtrsim 3.16\times10^{23}$ \& \newline $\lesssim 1.00\times10^{42}$ \\
         \hline
         $M$ & $4.40\times10^{5} \lesssim M \newline \lesssim 1.10\times10^{7}$ & -- & $2.55\times10^{5} \lesssim M \newline \lesssim 7.04\times10^{6}$\\
         \hline
         $f_*$ & $\gtrsim 0.05$ & -- & $\gtrsim 0.06$\\
         \hline
         $f_*$ \& $M$ & $\gtrsim 0.03$ \& \newline $\lesssim 8.53\times10^{8}$ & -- & $\gtrsim 0.02$ \& \newline $\lesssim 4.50\times10^{7}$\\
         \hline
    \end{tabular}
    \caption{\textbf{Key parameter constraints from SARAS3, HERA and the joint analysis.} Here the SARAS3 and HERA limits are taken from the respective papers. All the reported limits correspond to regions that are excluded with 68\% confidence. In the top two rows we show the type of signal targeted by each set of analysis and the corresponding redshifts. The joint analysis produces improved constraints on the radio and X-ray backgrounds while retaining the constraining power of SARAS3 on the star formation properties of early galaxies. $L_r$ is measured in W~Hz$^{-1}$~M$_\odot^{-1}$~yr at a reference frequency of $150$~MHz, $L_X$ in erg~s$^{-1}$~M$_\odot^{-1}$~yr, and is calculated by integrating X-ray spectral distribution of sources between 0.2 and 95~keV  assuming an X-ray SED consistent with that for X-ray binaries \protect\citep{Fragos_Xrays_2013}. The halo mass, $M$, is measured in solar masses. These constraints are derived from Kernel Density estimates~(KDE) of the marginal 1D and 2D posterior distributions.}
    \label{tab:numbers}
\end{table*}

We find comparable constraints on $f_*$ and minimum mass of star forming halos, $M$, as was found with SARAS3 alone \citep{Bevins_saras3_2022} when combining the data sets. Since these parameters predominantly affect the high redshift signal during the Cosmic Dawn they are better constrained by SARAS3, and we do not gain any information about star formation from HERA, which observes at $z\approx 8$ and $10$ during the Epoch of Reionization.
Marginalising over the radio and X-ray luminosities, we disfavour at 68\% confidence galaxies in which $\gtrsim2\%$ of gas is converted into stars and the minimum mass for star forming halos is $\lesssim 45$ million solar masses.

\subsection{Constraints on the radio and X-ray backgrounds}

\begin{figure*}
    \centering
    \includegraphics{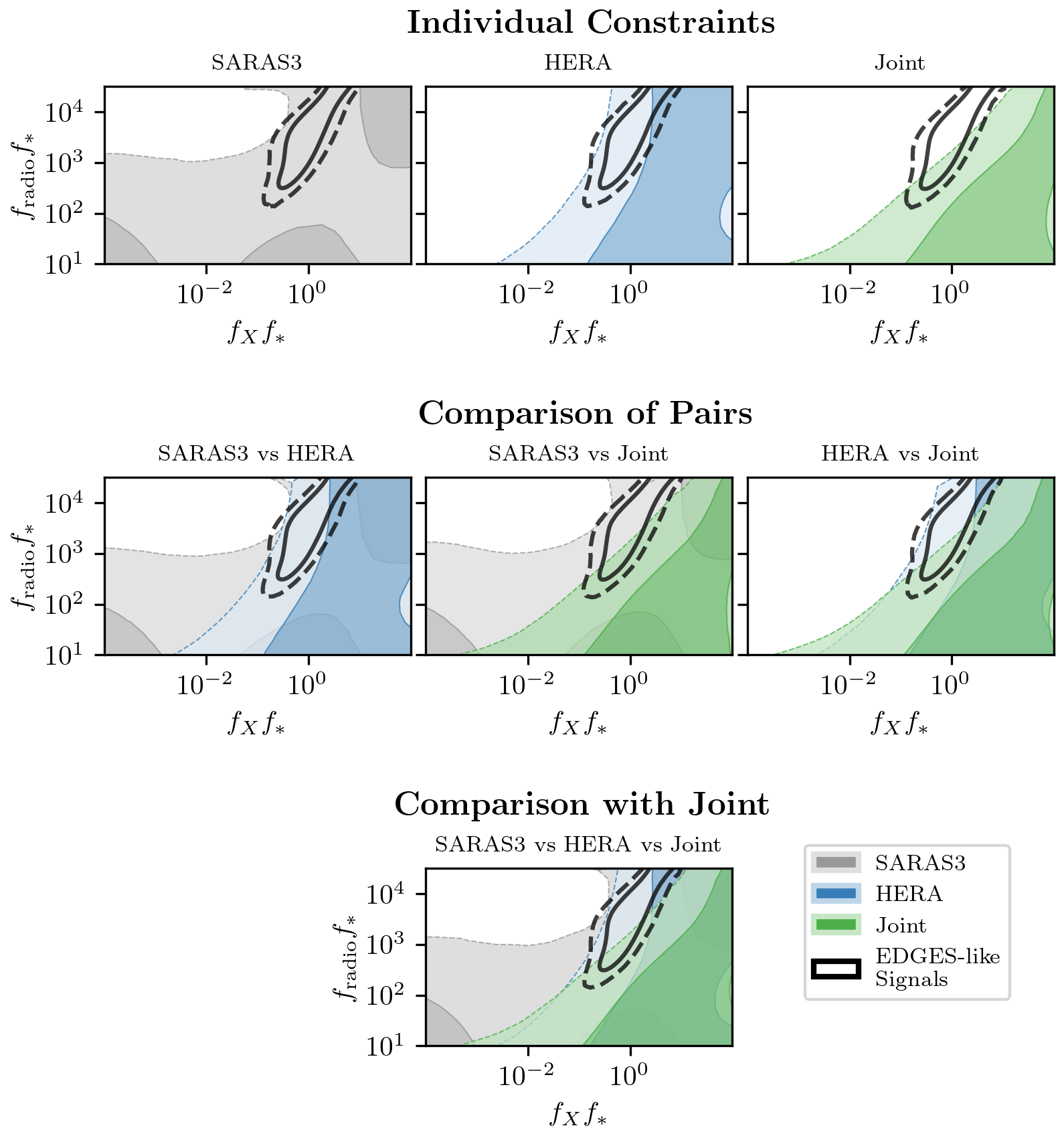}
    \caption{\textbf{Background constraints.} The figure shows constraints on the radio and X-ray backgrounds, parameterized by $f_* f_\mathrm{radio}$ and $f_* f_X$ respectively, from SARAS3 (grey), HERA (blue) and the joint analysis (green). The individual SARAS3 and HERA posteriors are based on the results presented in \protect\cite{Bevins_saras3_2022} and \protect\cite{HERA_2022b} respectively. We show each distribution individually on the top row,  overlaid pairs of distributions for comparison in the middle, and all three on the same figure in the bottom row. In all panels, we show 68\% and 95\% contours (black solid and dashed lines respectively) for physical signal models that have similar depths and central frequencies as the EDGES absorption feature as defined by the inequality in \protect\cite{Fialkov2019}. These physical EDGES-like models have previously been explored in the literature in \protect\cite{Fialkov2019, Reis2020}. We note that while individually both HERA and SARAS3 allow for astrophysically motivated signal models that could explain the depth of the EDGES feature, together they rule the corresponding parameter space out with approximately greater than $2\sigma$ confidence, although some EDGES-like signals are still viable. We stress again,  that the explored physical models cannot fully explain the shape of the EDGES signal.}
    \label{fig:background}
\end{figure*}

The product  $f_* f_\mathrm{radio}$ is proportional to the total radio background created by radio-luminous galaxies, and, equivalently,   $f_* f_X$ is a proxy for
the total X-ray background created by the early population of X-ray sources.
%The product of the star formation efficiency, $f_*$, with the radio production efficiency, $f_\mathrm{radio}$ and separately with the X-ray production efficiency, $f_X$, give proxies for the radio background and X-ray backgrounds respectively.
This is demonstrably true for our model parameterisation as the star formation rate is proportional to $f_*$ and the radio luminosity and X-ray luminosities per star formation rate are proportional to $f_\mathrm{radio}$ and $f_X$ respectively. Both X-ray and radio backgrounds are responsible for regulating the depth of the absorption feature, and they  can also be observed independently by other telescopes (e.g. observations of the unresolved X-ray background by Chandra \citep{Lehmer_Chandra_2012} and of the low-frequency radio background by ARCADE2/LWA \citep{fixsen11, dowell18}. In \cref{fig:background} we show constraints on the values of $f_*~f_X$ and $f_*~f_\mathrm{radio}$ achieved by SARAS3, HERA and the joint analysis. Since these combinations of the parameters regulate the absorption depth of the global 21-cm signal, we can also condition our prior on the astrophysical parameters to produce signals with the same central frequency and depth as the absorption feature found in the EDGES data \citep{Fialkov2019, Reis_lya_2021, Bevins_saras3_2022}. In each panel of  \cref{fig:background}, we show black contours corresponding to these EDGES-like physical signals. We see that while HERA and SARAS3 allow for combinations of $f_X~f_*$ and $f_\mathrm{radio}~f_*$ that could  partially explain EDGES, the combination of the two experiments, which produces a tighter constraint on the X-ray and radio luminosities of early galaxies, disfavours a large portion of the EDGES-like parameter space (i.e. most of the EDGES-like parameter space is beyond the 95\% contours of the joint constraints, while it is well within the 95\% contours for SARAS3 and HERA individually). This demonstrates further the power of combining different data sets. However, we note that the explored theoretical signals do not fit EDGES data well as none of them closely reproduces the flattened Gaussian-like feature found in the data \citep{EDGES}.

\subsection{Constraints on Phemenological Parameters}

\begin{figure*}
    \centering
    \includegraphics{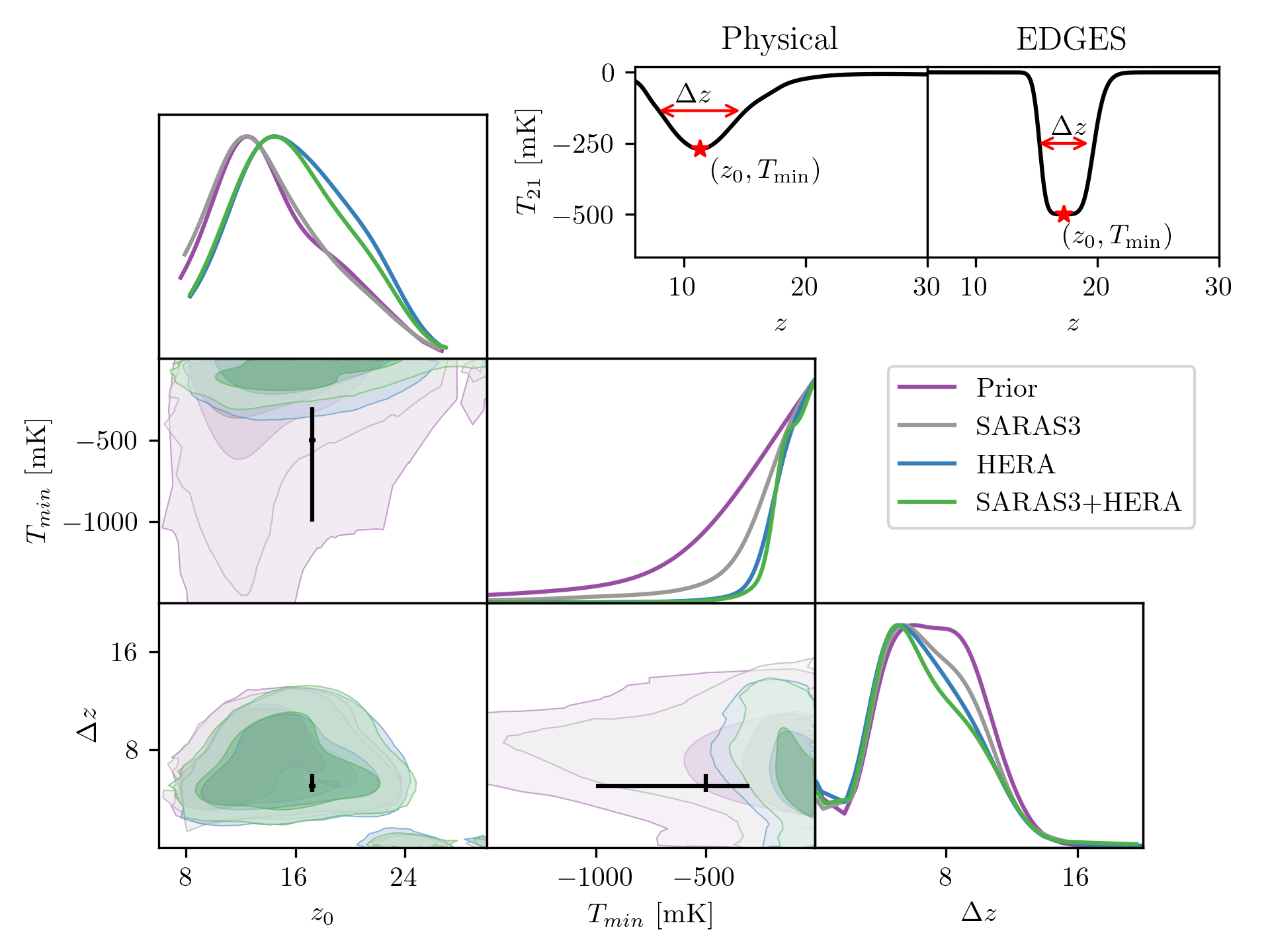}
    \caption{\textbf{Phenomenological constraints.} The triangular plots  shows the prior (purple) and posteriors (grey for SARAS3, blue for HERA, green for joint) of  the features of a typical global absorption signal: the central redshift, $z_0$, the corresponding minimum temperature, $T_\mathrm{min}$, and the width of the signal, $\Delta z$,  as is depicted in the top right corner. Darker shaded regions show $1\sigma$ constraints, lighter shaded regions show $2\sigma$ constraints. Overlaid on the posterior distributions are the 99\% confidence intervals, black crosses, reported for the corresponding phemenological parameterisation of the EDGES absorption feature in \protect\cite{EDGES}. Note that this is not the same as the physical EDGES-like distribution explored in \cref{fig:background}. %Note, for the physical signals these parameters are derived and they do not directly correspond to the EDGES $\delta z$, $T_\mathrm{min}$ and $z_0$. 
    We see that individually the experiments allow for  signals with depths that are consistent with EDGES. However, the combination of the two data sets disfavours these signals with greater than 2$\sigma$ significance. We do not disfavour signals with the same width or central frequency as EDGES, but note that the joint analysis indicates a preference for shallower and narrower signals with higher central redshifts  as can be seen in the 1D PDFs.}
    \label{fig:tmin}
\end{figure*}

We explore the structure of the global 21-cm signals which are consistent with the data from SARAS3, HERA and the two sets together. In order to do this we look at the minimum temperature, $T_\mathrm{min}$, the  corresponding central redshift, $z_0$, and an approximate full width at half max, $\Delta z$, of each absorption trough as defined in the top right of \cref{fig:tmin}. More specifically, for each parameter set $\theta_{21}$ (either from the prior parameter range, or from the posterior ranges consistent with SARAS3, HERA, and the joint analysis) we generate a global signal using the  neural network emulator \textsc{globalemu} \citep{globalemu}. We then measure the values of $T_\mathrm{min}$, $z_0$, and  $\Delta z$ for each signal producing probability distributions for each parameter corresponding to the constraints from each data set and showing the extent of the prior. We compare these distributions with the values used to parameterize the phenomenological EDGES flattened Gaussian signal with the EDGES 99\% confidence ranges shown by the black crosses in the corner plot in \cref{fig:tmin}.

When considered individually, the SARAS3 and HERA experiments allow for astrophysically motivated sky-averaged 21-cm signals that have a minimum temperature, location and width in agreement with the EDGES detection, while the joint analysis rules out models with a depth that is consistent with EDGES at greater than a $2\sigma$ significance. The joint analysis has a  preference for shallower (lower values of $|T_\mathrm{min}|$) and narrower signals with higher central frequencies, as can be seen in the corresponding 1D PDFs, which is driven largely by the HERA constraints.

\subsection{The impact of SARAS2} \label{sec:saras2}

Previous analysis of the SARAS2 data revealed some weak constraints, most notably in the plane of $L_X - L_{r}$ in agreement with HERA and SARAS3, on the properties of galaxies in the infant Universe \citep{Bevins_saras2_2022}. SARAS2 is at much lower redshifts than SARAS3 but overlaps with the redshifts probed by HERA having recorded observations in the band $z\approx7-12$. The data is contaminated by a sinusoidal systematic and a number of different models were fitted to this feature. The systematic models correspond to a signal introduced prior to the antenna possibly from ground emission or some unknown component of the foreground, and separately a signal introduced in the system electronics potentially from cable reflections. The sinusoidal systematic was fitted alongside a signal model generated with \textsc{globalemu} and a foreground model that is conditioned to be smooth preventing it fitting out non-smooth systematics or signals in the data \citep{maxsmooth}. Here we take the best fitting model, with a systematic from ground emission or a non-smooth component from the foreground, with the highest evidence from the original analysis \citep{Bevins_saras2_2022} and combine the corresponding constraints on the astrophysical parameters $V_c$, $f_*$, $f_X$, $f_\mathrm{radio}$ and $\tau$ with the joint constraints from HERA and SARAS3 to assess the impact.

\begin{figure*}
    \centering
    \includegraphics{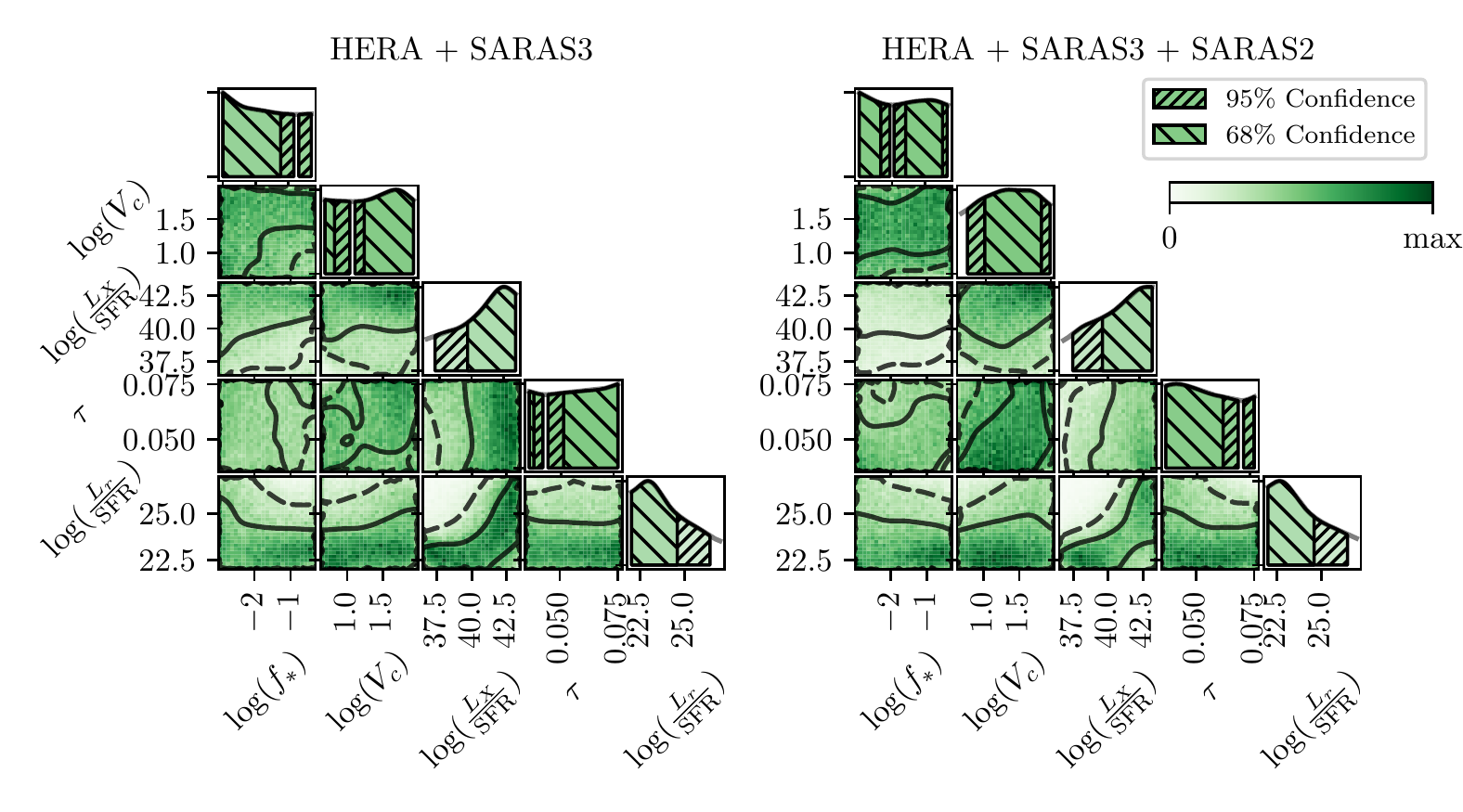}
    \caption{\textbf{The impact of SARAS2 data on the astrophysical constraints.} We show the joint posterior distributions for HERA and SARAS3 on the left panel (identical to Figure \ref{fig:posterior}, but shown here for comparison) and for HERA, SARAS3 and SARAS2 on the right panel. SARAS2 covers the band $z\approx 7 - 12$ and therefore has some overlap with HERA but not with SARAS3. The addition of SARAS2 to the joint analysis washes out the constraint on star formation properties, $V_c$ and $f_*$, because it leads to increased uncertainty in the structure of the signals at high redshifts. However, we still see a consistent disfavouring of a population of radio galaxies with high radio and low X-ray luminosities. The one dimensional posteriors for $\tau$ appear to be in disagreement, however, we note that these are basically flat. We exclude SARAS2 from our main results in the text because of uncertainty in the modelling of systematics in the data.}
    \label{fig:other_global}
\end{figure*}

As can be seen in \cref{fig:other_global}, the addition of SARAS2 to our analysis washes out the constraint in the plane $f_* - V_c$. One possible explanation for this is that the addition of SARAS2, while constraining the properties that affect the signal at low redshifts, increases the envelope of possible models at higher redshifts, where star formation is more important, that are plausible even given the constraints from the SARAS3 data. Despite this, we note that we maintain the constraint in the plane $L_X - L_r$ when we add SARAS2 into our analysis.

We can quantify the impact of SARAS2 on our analysis by looking at the percentage of the astrophysical prior volume which is consistent with the different combinations of the three different data sets. To calculate this percentage, we use \textsc{margarine} to calculate the marginal Kullback-Leibler~(KL) divergence \citep{kullback_information_1951}, $\mathcal{D}$, between the flat prior on the five astrophysical parameters in the set $\theta_{21}$ and the corresponding posteriors. The KL divergence gives a measure of the information gained when moving from the prior to the posterior and is related to the percentage via
\begin{equation}
    \% = 100\times\exp(-\mathcal{D}) \approx 100 \times \frac{V_\mathcal{P}}{V_\pi},
\end{equation}
where $V_\pi$ is the prior volume and $V_\mathcal{P}$ is the posterior volume. This quantity is useful as it quantifies the constraining power of the different data sets in all five dimensions, including correlations that may not be visible in the one and two-dimensional projections used to produce the corner plots in this paper and in the literature. 

\begin{figure*}
    \centering
    \includegraphics{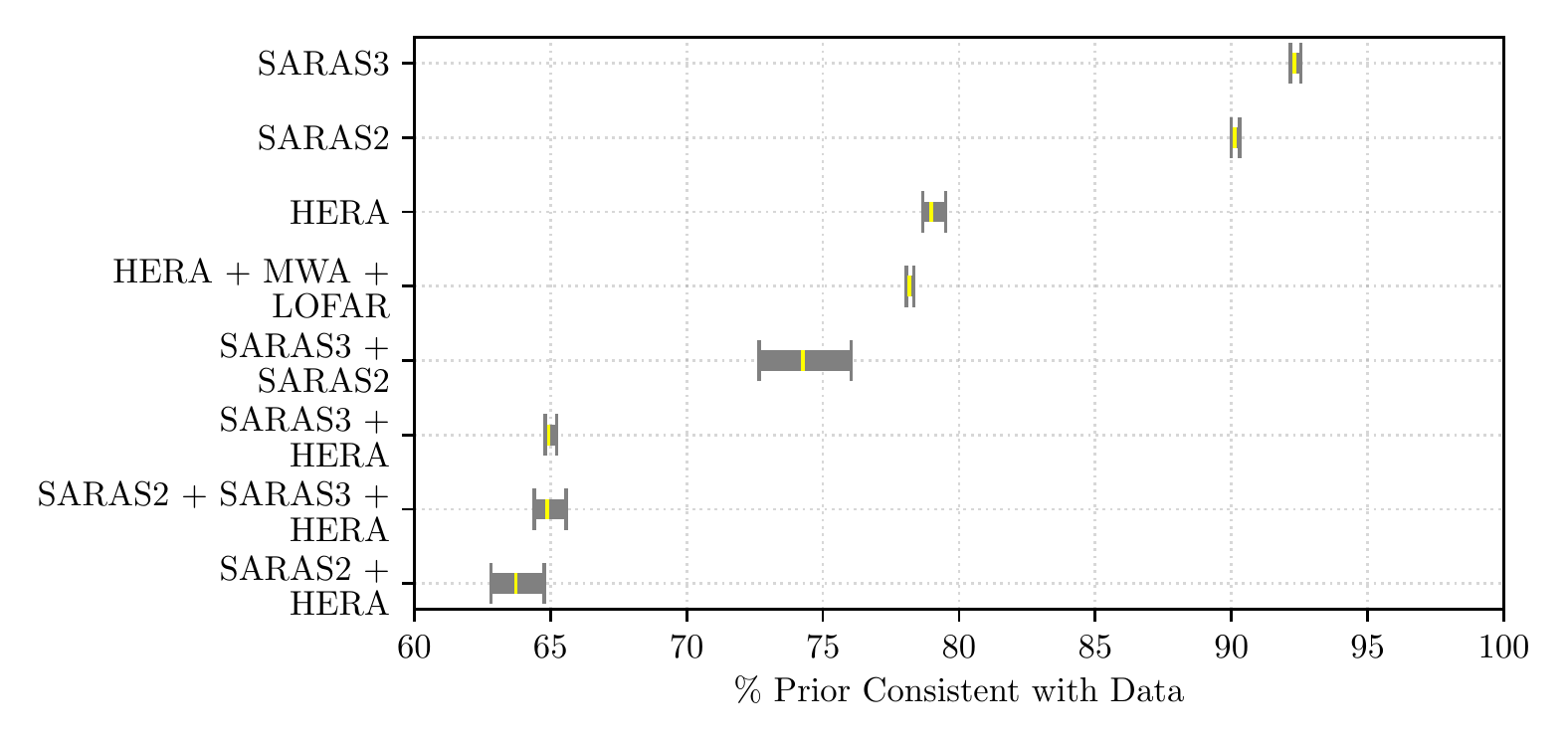}
    \caption{\textbf{Constraining power of different data sets.} The percentage of the wide astrophysical parameter prior that is found to be consistent with the different data sets and different combinations of data sets explored in this work. A lower value indicates a better set of constraints, although a difference of a few percent does not necessarily translate into significant differences in the parameter constraints as can be seen when comparing the results from HERA and HERA + LOFAR + MWA.}
    \label{fig:volume_contraction}
\end{figure*}

We show in \cref{fig:volume_contraction} the percentage of the astrophysical parameter prior volume that is consistent with different combinations of the data sets discussed in this work~(including additional interferometric measurements of the power spectrum discussed in \cref{sec:mwa_lofar}). We see that the combination of either or both of the SARAS data sets with HERA lead to a percentage consistency with the data of $\approx 63 - 65\%$ and this is likely dominated by HERA. Individually, HERA allows for $\approx 80\%$ of the astrophysical parameter space, SARAS2 for $\approx90\%$ and SARAS3 for $\approx92\%$. Due to the uncertainty in the modelling of the systematics in the SARAS2 analysis, we leave SARAS2 out of the main results.

\subsection{The impact of MWA and LOFAR} \label{sec:mwa_lofar}

\begin{figure*}
    \centering
    \includegraphics{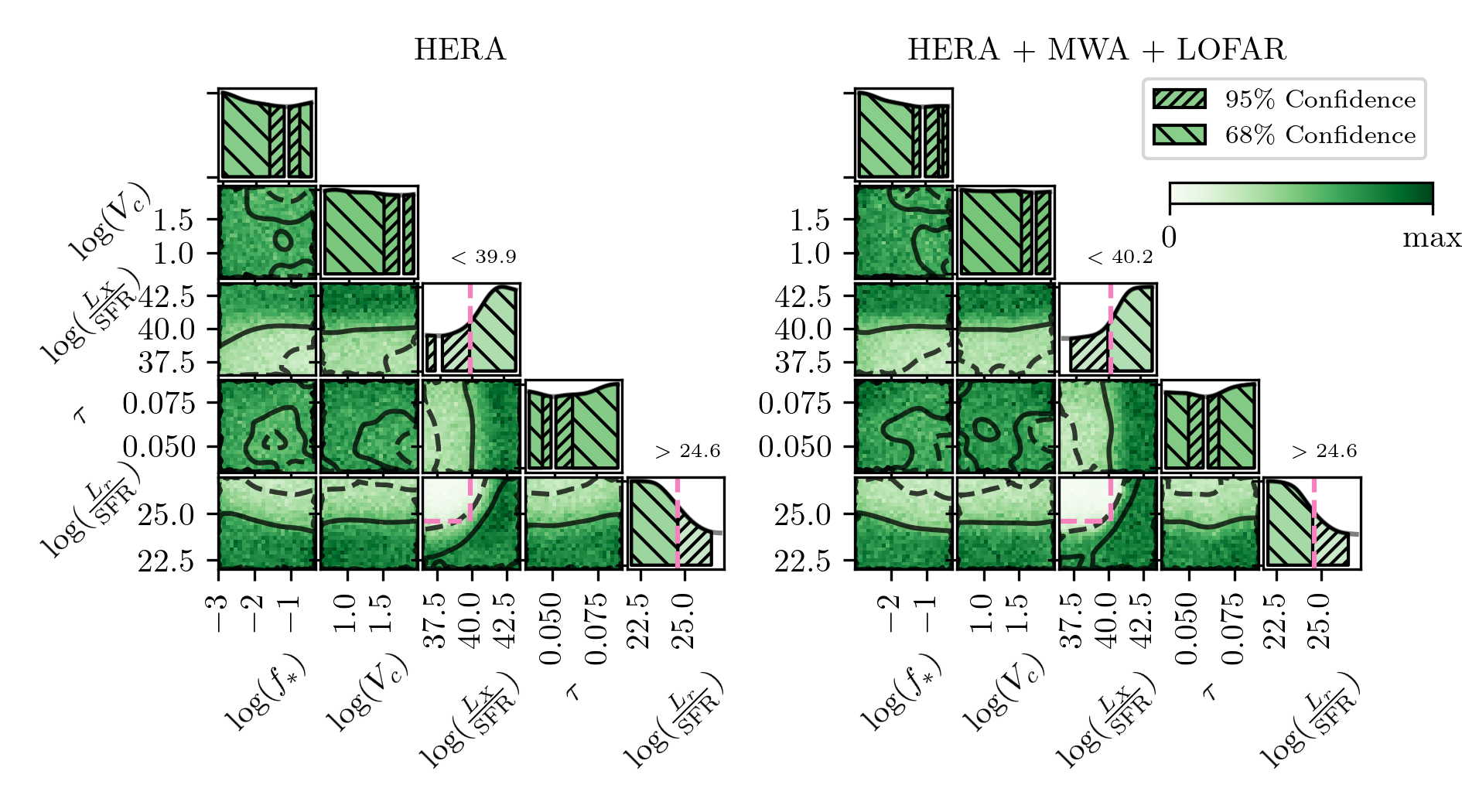}
    \caption{\textbf{The impact of MWA and LOFAR on the parameter constraints.} Projected posterior distribution functions (PDFs) for the 5 simulation parameters, obtained by assuming flat priors and combining different observations: HERA alone \protect\citep[left, as in][]{HERA_2022b} and LOFAR, HERA and MWA (right). Solid (dashed) lines correspond to regions containing the highest 68\% (95\%) probability. We see that HERA constraints are not significantly improved by adding the published limits on the power spectrum from other interferometers.}
    \label{fig:oi}
\end{figure*}

In \cref{fig:oi}, we show the projected posteriors derived using HERA data alone (left panel) and HERA, MWA and LOFAR together (right panel). We note that the constraints from the different interferometers are all at low redshifts between $z\approx 6 - 10$ and varying wavenumbers or angular scales. These are detailed in \cref{tab:interferometers} along with the constraints from the individual experiments on key parameters.

We derived parameter constraints from the MWA and LOFAR data using the approach taken in the orginal HERA analysis. Specifically, we take the measured upper limits, the mean power spectrum and uncertainty, and treat it as a measurement of cosmological signal plus systematics. As in HERA \citep{HERA_2022b} we take this uncertainty to be Gaussian and marginalize over a uniform prior on the systematics, yielding the likelihood

\begin{equation}
    \mathcal{L}(\theta_{21}) = \prod_i^{N_d}\frac{1}{2}\bigg(1 - \mathrm{erf}\bigg[\frac{d_i - m_i(\theta_{21})}{\sqrt{2\sigma_i}}\bigg]\bigg), 
\end{equation}
where $N_d$ represents the number of data points, $d_i$ and $\sigma_i$ correspond to the mean and an error term which includes a contribution from the standard deviation in each data point and an emulator error, and $m_i(\theta_{21})$ is the model prediction for that redshift and wave number. Thus a model prediction $m\gg d$ gives $\mathcal{L}\approx0$ while $m\ll d$ gives a constant. This likelihood is effectively a step function that disfavours models above a given amplitude. A full discussion of its derivation can be found in \cite{HERA_2022b}.

We performed this joint analysis using a full analytic likelihood approach~(independent of \textsc{margarine}) since there are no nuisance parameters describing the systematics. We find that each of the experiments disfavours individually similar regions of the $L_r - L_X$ plane. However, the joint analysis does not improve the results derived from HERA data alone~(as we summarise in \cref{tab:interferometers} and is further illustrated in \cref{fig:volume_contraction}) which motivates our decision to only use HERA in our main results with SARAS3.

\begin{table*}
    \centering
    \begin{tabular}{|p{2cm}|p{3cm}|p{3cm}|p{3cm}|p{3cm}|}
        \hline
         & HERA  & LOFAR & MWA & LOFAR + MWA + HERA \\
         \hline
         $z$ & $\approx 8$ \&  $\approx 10$ & $9.1$ \&  $ 9.3 - 10.6$ & $6.5 - 8.7$ & Discrete and continuous ranges of $z$ between $6.5 - 10.6$ \\
         \hline
         $k$ [h~Mpc$^{-1}$]& $0.128 - 0.960$  & $0.075 - 0.432$ \&  $0.053$& $0.070-3.000$ & Discrete and continuous ranges of $k$ \\
         \hline
         \hline
         $L_{r}/\mathrm{SFR}$ & $\gtrsim4.00\times10^{24}$ & $\gtrsim 1.20 \times10^{25}$ & $\gtrsim1.58\times10^{25}$ & $\gtrsim4.00\times10^{24}$ \\
         \hline
         $L_{X}/\mathrm{SFR}$ & $\lesssim7.60\times10^{39}$ & $\lesssim 8.70 \times10^{38}$ & $\lesssim1.16\times10^{39}$ & $\lesssim 1.58 \times10^{40}$\\
         \hline
         $L_{r}/\mathrm{SFR}$ \&  $L_{X}/\mathrm{SFR}$ & $\gtrsim 4.00\times10^{24}$ \&  $\lesssim 7.60\times10^{39}$ & $\gtrsim 3.16\times10^{25}$ \&  $\lesssim 1.00\times10^{40}$ & $\gtrsim 1.00\times10^{25}$ \&  $\lesssim 1.00\times10^{40}$ & $\gtrsim4.00\times10^{24}$ \&  $\lesssim1.58 \times10^{40}$ \\
         \hline
    \end{tabular}
    \caption{\textbf{Constraints from interferometers.} The table shows the various constraints on the radio and X-ray luminosities for HERA, MWA, LOFAR and the combination of all three along with their respective wavenumbers and redshift ranges. The joint analysis only marginally improves our understanding of the infant universe.}
    \label{tab:interferometers}
\end{table*}

\section{Conclusion}\label{sec:conclusions}

Through a combination of constraints on fluctuations  and the sky-averaged 21-cm signal of neutral hydrogen, we have improved our understanding of the first galaxies that formed in the infant Universe between 200 and 700 million years after the Big Bang. This is the first time the data from the two different 21-cm probes have been combined to derive constraints on the astrophysical properties of the early galaxies. Even though the existing constraints are weak and model dependent, we develop novel methodology and outline the approach, which will become increasingly more useful as the  next-generation experiments  deliver stronger observational constraints.

Considering a wide space of plausible astrophysical  models including high-redshift sources of ultraviolet, X-ray and radio photons which affect the 21-cm signal, we calculate corresponding sky-averaged spectra as well as the power spectra of fluctuations. Using an upper limit on the fluctuations from the HERA interferometer and non-detection of the global 21-cm signal by the SARAS3 radiometer, we find that only $64.9^{+0.3}_{-0.1}$\% of the explored theoretical parameter space is consistent with the joint SARAS3 and HERA constraint, which is a significant improvement over the individual values of $92.3^{+0.3}_{-0.1}$\% and $78.7\pm 0.2$\% respectively.

Using the newly developed methodology we place the tightest current constraints on the properties of cosmic gas, such as the spin temperature of the 21-cm hydrogen line (closely related, but not equal, to the gas kinetic temperature) and the radio background temperature, $T_{r}$, as well as  on the radio and X-ray luminosities of the first galaxies  disfavouring at 68\% confidence galaxies that are approximately 32 times more efficient radio emitters than present galaxies and simultaneously are less than 33 times bright in the X-ray band. This work reports an increased degree of confidence over a wider range of redshifts than previous works which typically extrapolate outside the redshifts targeted by individual experiments, while here we interpolated between the observations of SARAS3 at $z = 15-25$ and the HERA limits at lower redshifts $z\sim8$ and 10. 

In this work we also considered the addition of  interferometric data sets from  MWA and LOFAR to our analysis, which, owing to the current weaker limits reported by these experiments, led to a negligible  improvement in the results. Similarly to HERA, these experiments probe the physics of the EoR covering  a similar redshift range to HERA  (between $z\approx 6 - 10$), with the current HERA data providing the tightest constraints on our models. Of the current global or sky-averaged 21-cm experiments only SARAS2, SARAS3 and EDGES were able to place limits on the astrophysics of the infant Universe. Our main focus is on the SARAS3 limits (although we also consider SARAS2), as there are concerns surrounding the cosmological nature of the signal reported in EDGES and a degree of uncertainty in the modelling of systematics in the SARAS2 data. We note that the SARAS2 data covers a similar redshift range as the HERA data and based on our  analysis does not lead to a significant improvement in our constraints.

The new methodology developed in this paper will allow for synergies between the upcoming observations, e.g. of the power spectrum from cosmic dawn measured by the NenuFAR \citep{Zarka_nenuFar_2018}, HERA, LOFAR or the SKA \citep{Mellema_SKA_2013}, as well as the measurements of the global signal by the wide-band REACH experiment covering the redshift range $z\approx 7 - 28$ \citep{REACH},  PRIZM \citep{Philip_prizm_2019}, MIST \citep{MIST} and missions to the moon \citep{Burns_2021} among others. 

\section*{Acknowledgements}

HTJB acknowledges the support of the Science and Technology Facilities Council (STFC) through grant number ST/T505997/1. IAC acknowledges support from the Institute of Astronomy Summer Internship Program at the University of Cambridge. WJH and AF were supported by Royal Society University Research Fellowships. EdLA was supported by the STFC through the Ernest Rutherford Fellowship. RB acknowledges the support of the Israel Science Foundation (grant No. 2359/20), The Ambrose Monell Foundation and the Institute for Advanced Study.

\section*{Data Availability}

The SARAS3 and SARAS2 data are available upon reasonable request to SS. The HERA data is publicly available at  \url{https://reionization.org}. The MWA and LOFAR data were provided to IAC through private correspondence with the relevant authors.

\textsc{margarine} is available at \url{https://github.com/htjb/margarine}. The semi-analytic simulations of the 21-cm signal discussed in \cref{sec:method} are not publicly available. \textsc{globalemu} is available at \url{https://github.com/htjb/globalemu} and the power spectrum emulators used are not publicly available but are built with \textsc{sklearn} (\url{https://scikit-learn.org/stable/install.html}). \textsc{polychord} is available at \url{https://github.com/PolyChord/PolyChordLite}. The graphs in the paper were made with \textsc{anesthetic} which is available at \url{https://github.com/handley-lab/anesthetic} and \textsc{fgivenx} which is available at \url{https://github.com/handley-lab/fgivenx}.


\bibliography{example}% Produces the bibliography via BibTeX.

\begin{thebibliography}{}
\makeatletter
\relax
\def\mn@urlcharsother{\let\do\@makeother \do\$\do\&\do\#\do\^\do\_\do\%\do\~}
\def\mn@doi{\begingroup\mn@urlcharsother \@ifnextchar [ {\mn@doi@}
  {\mn@doi@[]}}
\def\mn@doi@[#1]#2{\def\@tempa{#1}\ifx\@tempa\@empty \href
  {http://dx.doi.org/#2} {doi:#2}\else \href {http://dx.doi.org/#2} {#1}\fi
  \endgroup}
\def\mn@eprint#1#2{\mn@eprint@#1:#2::\@nil}
\def\mn@eprint@arXiv#1{\href {http://arxiv.org/abs/#1} {{\tt arXiv:#1}}}
\def\mn@eprint@dblp#1{\href {http://dblp.uni-trier.de/rec/bibtex/#1.xml}
  {dblp:#1}}
\def\mn@eprint@#1:#2:#3:#4\@nil{\def\@tempa {#1}\def\@tempb {#2}\def\@tempc
  {#3}\ifx \@tempc \@empty \let \@tempc \@tempb \let \@tempb \@tempa \fi \ifx
  \@tempb \@empty \def\@tempb {arXiv}\fi \@ifundefined
  {mn@eprint@\@tempb}{\@tempb:\@tempc}{\expandafter \expandafter \csname
  mn@eprint@\@tempb\endcsname \expandafter{\@tempc}}}

\bibitem[\protect\citeauthoryear{{Abdurashidova} et~al.,}{{Abdurashidova}
  et~al.}{2022a}]{HERA_2022b}
{Abdurashidova} Z.,  et~al., 2022a, \mn@doi [\apj] {10.3847/1538-4357/ac2ffc},
  \href {https://ui.adsabs.harvard.edu/abs/2022ApJ...924...51A} {924, 51}

\bibitem[\protect\citeauthoryear{{Abdurashidova} et~al.,}{{Abdurashidova}
  et~al.}{2022b}]{HERA_2022a}
{Abdurashidova} Z.,  et~al., 2022b, \mn@doi [\apj] {10.3847/1538-4357/ac1c78},
  \href {https://ui.adsabs.harvard.edu/abs/2022ApJ...925..221A} {925, 221}

\bibitem[\protect\citeauthoryear{{Acharya}, {Dhandha}  \& {Chluba}}{{Acharya}
  et~al.}{2022}]{Acharya_2022}
{Acharya} S.~K.,  {Dhandha} J.,   {Chluba} J.,  2022, \mn@doi [\mnras]
  {10.1093/mnras/stac2739}, \href
  {https://ui.adsabs.harvard.edu/abs/2022MNRAS.517.2454A} {517, 2454}

\bibitem[\protect\citeauthoryear{{Acharya}, {Cyr}  \& {Chluba}}{{Acharya}
  et~al.}{2023}]{Acharya_2023}
{Acharya} S.~K.,  {Cyr} B.,   {Chluba} J.,  2023, \mn@doi [\mnras]
  {10.1093/mnras/stad1540}, \href
  {https://ui.adsabs.harvard.edu/abs/2023MNRAS.523.1908A} {523, 1908}

\bibitem[\protect\citeauthoryear{{Ashton} et~al.,}{{Ashton}
  et~al.}{2022}]{Ashton_ns_review_2022}
{Ashton} G.,  et~al., 2022, arXiv e-prints, \href
  {https://ui.adsabs.harvard.edu/abs/2022arXiv220515570A} {p. arXiv:2205.15570}

\bibitem[\protect\citeauthoryear{{Barkana}}{{Barkana}}{2016}]{Barkana_review_2016}
{Barkana} R.,  2016, \mn@doi [\physrep] {10.1016/j.physrep.2016.06.006}, \href
  {https://ui.adsabs.harvard.edu/abs/2016PhR...645....1B} {645, 1}

\bibitem[\protect\citeauthoryear{{Barkana}}{{Barkana}}{2018}]{BarkanaDM2018}
{Barkana} R.,  2018, \mn@doi [Nature] {10.1038/nature25791}, \href
  {https://ui.adsabs.harvard.edu/abs/2018Natur.555...71B} {555, 71}

\bibitem[\protect\citeauthoryear{{Barkana}, {Outmezguine}, {Redigol}  \&
  {Volansky}}{{Barkana} et~al.}{2018}]{Barkana2018}
{Barkana} R.,  {Outmezguine} N.~J.,  {Redigol} D.,   {Volansky} T.,  2018,
  \mn@doi [Phys. Rev.~D] {10.1103/PhysRevD.98.103005}, 98, 103005

\bibitem[\protect\citeauthoryear{{Berlin}, {Hooper}, {Krnjaic}  \&
  {McDermott}}{{Berlin} et~al.}{2018}]{Berlin2018}
{Berlin} A.,  {Hooper} D.,  {Krnjaic} G.,   {McDermott} S.~D.,  2018, \mn@doi
  [prl] {10.1103/PhysRevLett.121.011102}, 121, 011102

\bibitem[\protect\citeauthoryear{{Bernardi} et~al.,}{{Bernardi}
  et~al.}{2016}]{Bernardi_LEDA_2016}
{Bernardi} G.,  et~al., 2016, \mn@doi [\mnras] {10.1093/mnras/stw1499}, \href
  {https://ui.adsabs.harvard.edu/abs/2016MNRAS.461.2847B} {461, 2847}

\bibitem[\protect\citeauthoryear{{Bevins}, {Handley}, {Fialkov}, {de Lera
  Acedo}, {Greenhill}  \& {Price}}{{Bevins} et~al.}{2021a}]{maxsmooth}
{Bevins} H.~T.~J.,  {Handley} W.~J.,  {Fialkov} A.,  {de Lera Acedo} E.,
  {Greenhill} L.~J.,   {Price} D.~C.,  2021a, \mn@doi [\mnras]
  {10.1093/mnras/stab152}, \href
  {https://ui.adsabs.harvard.edu/abs/2021MNRAS.502.4405B} {502, 4405}

\bibitem[\protect\citeauthoryear{{Bevins}, {Handley}, {Fialkov}, {de Lera
  Acedo}  \& {Javid}}{{Bevins} et~al.}{2021b}]{globalemu}
{Bevins} H.~T.~J.,  {Handley} W.~J.,  {Fialkov} A.,  {de Lera Acedo} E.,
  {Javid} K.,  2021b, \mn@doi [\mnras] {10.1093/mnras/stab2737}, \href
  {https://ui.adsabs.harvard.edu/abs/2021MNRAS.508.2923B} {508, 2923}

\bibitem[\protect\citeauthoryear{{Bevins}, {Handley}, {Lemos}, {Sims}, {de Lera
  Acedo}, {Fialkov}  \& {Alsing}}{{Bevins} et~al.}{2022a}]{margarine}
{Bevins} H. T.~J.,  {Handley} W.~J.,  {Lemos} P.,  {Sims} P.~H.,  {de Lera
  Acedo} E.,  {Fialkov} A.,   {Alsing} J.,  2022a, arXiv e-prints, \href
  {https://ui.adsabs.harvard.edu/abs/2022arXiv220512841B} {p. arXiv:2205.12841}

\bibitem[\protect\citeauthoryear{{Bevins}, {Handley}, {Lemos}, {Sims}, {de Lera
  Acedo}  \& {Fialkov}}{{Bevins} et~al.}{2022b}]{margarine2}
{Bevins} H.,  {Handley} W.,  {Lemos} P.,  {Sims} P.,  {de Lera Acedo} E.,
  {Fialkov} A.,  2022b, arXiv e-prints, \href
  {https://ui.adsabs.harvard.edu/abs/2022arXiv220711457B} {p. arXiv:2207.11457}

\bibitem[\protect\citeauthoryear{{Bevins}, {Fialkov}, {de Lera Acedo},
  {Handley}, {Singh}, {Subrahmanyan}  \& {Barkana}}{{Bevins}
  et~al.}{2022c}]{Bevins_saras3_2022}
{Bevins} H.~T.~J.,  {Fialkov} A.,  {de Lera Acedo} E.,  {Handley} W.~J.,
  {Singh} S.,  {Subrahmanyan} R.,   {Barkana} R.,  2022c, \mn@doi [Nature
  Astronomy] {10.1038/s41550-022-01825-6}, \href
  {https://ui.adsabs.harvard.edu/abs/2022NatAs...6.1473B} {6, 1473}

\bibitem[\protect\citeauthoryear{{Bevins}, {de Lera Acedo}, {Fialkov},
  {Handley}, {Singh}, {Subrahmanyan}  \& {Barkana}}{{Bevins}
  et~al.}{2022d}]{Bevins_saras2_2022}
{Bevins} H.~T.~J.,  {de Lera Acedo} E.,  {Fialkov} A.,  {Handley} W.~J.,
  {Singh} S.,  {Subrahmanyan} R.,   {Barkana} R.,  2022d, \mn@doi [\mnras]
  {10.1093/mnras/stac1158}, \href
  {https://ui.adsabs.harvard.edu/abs/2022MNRAS.513.4507B} {513, 4507}

\bibitem[\protect\citeauthoryear{{Bond}}{{Bond}}{1981}]{Bond_pop3_1981}
{Bond} H.~E.,  1981, \mn@doi [\apj] {10.1086/159186}, \href
  {https://ui.adsabs.harvard.edu/abs/1981ApJ...248..606B} {248, 606}

\bibitem[\protect\citeauthoryear{Bowman, Rogers, Monsalve, Mozdzen  \&
  Mahesh}{Bowman et~al.}{2018}]{EDGES}
Bowman J.~D.,  Rogers A. E.~E.,  Monsalve R.~A.,  Mozdzen T.~J.,   Mahesh N.,
  2018, \mn@doi [Nature] {10.1038/nature25792}, 555, 67

\bibitem[\protect\citeauthoryear{{Bradley}, {Tauscher}, {Rapetti}  \&
  {Burns}}{{Bradley} et~al.}{2019}]{Bradley_EDGES_2019}
{Bradley} R.~F.,  {Tauscher} K.,  {Rapetti} D.,   {Burns} J.~O.,  2019, \mn@doi
  [ApJ] {10.3847/1538-4357/ab0d8b}, \href
  {https://ui.adsabs.harvard.edu/abs/2019ApJ...874..153B} {874, 153}

\bibitem[\protect\citeauthoryear{Bromm \& Larson}{Bromm \&
  Larson}{2004}]{Bromm_pop3_2004}
Bromm V.,  Larson R.~B.,  2004, \mn@doi [Annual Review of Astronomy and
  Astrophysics] {10.1146/annurev.astro.42.053102.134034}, 42, 79

\bibitem[\protect\citeauthoryear{{Burns} et~al.,}{{Burns}
  et~al.}{2021}]{Burns_2021}
{Burns} J.,  et~al., 2021, arXiv e-prints, \href
  {https://ui.adsabs.harvard.edu/abs/2021arXiv210305085B} {p. arXiv:2103.05085}

\bibitem[\protect\citeauthoryear{{Caputo}, {Liu}, {Mishra-Sharma}, {Pospelov}
  \& {Ruderman}}{{Caputo} et~al.}{2023}]{Caputo_2023}
{Caputo} A.,  {Liu} H.,  {Mishra-Sharma} S.,  {Pospelov} M.,   {Ruderman}
  J.~T.,  2023, \mn@doi [\prd] {10.1103/PhysRevD.107.123033}, \href
  {https://ui.adsabs.harvard.edu/abs/2023PhRvD.107l3033C} {107, 123033}

\bibitem[\protect\citeauthoryear{{Chuzhoy} \& {Shapiro}}{{Chuzhoy} \&
  {Shapiro}}{2007}]{Chuzhoy2007}
{Chuzhoy} L.,  {Shapiro} P.~R.,  2007, \mn@doi [\apj] {10.1086/510146}, \href
  {https://ui.adsabs.harvard.edu/abs/2007ApJ...655..843C} {655, 843}

\bibitem[\protect\citeauthoryear{{Cohen}, {Fialkov}, {Barkana}  \&
  {Lotem}}{{Cohen} et~al.}{2017}]{Cohen_global_2017}
{Cohen} A.,  {Fialkov} A.,  {Barkana} R.,   {Lotem} M.,  2017, \mn@doi [\mnras]
  {10.1093/mnras/stx2065}, \href
  {https://ui.adsabs.harvard.edu/abs/2017MNRAS.472.1915C} {472, 1915}

\bibitem[\protect\citeauthoryear{{Cohen}, {Fialkov}  \& {Barkana}}{{Cohen}
  et~al.}{2018}]{Cohen_power_2018}
{Cohen} A.,  {Fialkov} A.,   {Barkana} R.,  2018, \mn@doi [\mnras]
  {10.1093/mnras/sty1094}, \href
  {https://ui.adsabs.harvard.edu/abs/2018MNRAS.478.2193C} {478, 2193}

\bibitem[\protect\citeauthoryear{DeBoer et~al.,}{DeBoer
  et~al.}{2017}]{DeBoer_HERA_2017}
DeBoer D.,  et~al., 2017, \mn@doi [\pasp] {10.1088/1538-3873/129/974/045001},
  129, 45001

\bibitem[\protect\citeauthoryear{{Dowell} \& {Taylor}}{{Dowell} \&
  {Taylor}}{2018}]{dowell18}
{Dowell} J.,  {Taylor} G.~B.,  2018, \mn@doi [\apjl]
  {10.3847/2041-8213/aabf86}, \href
  {https://ui.adsabs.harvard.edu/abs/2018ApJ...858L...9D} {858, L9}

\bibitem[\protect\citeauthoryear{{Eastwood} et~al.,}{{Eastwood}
  et~al.}{2019}]{LWA_PS_2019}
{Eastwood} M.~W.,  et~al., 2019, \mn@doi [\aj] {10.3847/1538-3881/ab2629},
  \href {https://ui.adsabs.harvard.edu/abs/2019AJ....158...84E} {158, 84}

\bibitem[\protect\citeauthoryear{{Ewall-Wice}, {Chang}, {Lazio}, {Dor{\'e}},
  {Seiffert}  \& {Monsalve}}{{Ewall-Wice} et~al.}{2018}]{Ewall2018}
{Ewall-Wice} A.,  {Chang} T.~C.,  {Lazio} J.,  {Dor{\'e}} O.,  {Seiffert} M.,
  {Monsalve} R.~A.,  2018, \mn@doi [ApJ] {10.3847/1538-4357/aae51d}, 868, 63

\bibitem[\protect\citeauthoryear{{Feng} \& {Holder}}{{Feng} \&
  {Holder}}{2018}]{Feng2018}
{Feng} C.,  {Holder} G.,  2018, \mn@doi [\apjl] {10.3847/2041-8213/aac0fe},
  858, L17

\bibitem[\protect\citeauthoryear{{Fialkov} \& {Barkana}}{{Fialkov} \&
  {Barkana}}{2014}]{Fialkov_rich_2014}
{Fialkov} A.,  {Barkana} R.,  2014, \mn@doi [\mnras] {10.1093/mnras/stu1744},
  \href {https://ui.adsabs.harvard.edu/abs/2014MNRAS.445..213F} {445, 213}

\bibitem[\protect\citeauthoryear{Fialkov \& Barkana}{Fialkov \&
  Barkana}{2019}]{Fialkov2019}
Fialkov A.,  Barkana R.,  2019, \mn@doi [Monthly Notices of the Royal
  Astronomical Society] {10.1093/mnras/stz873}, 486, 1763

\bibitem[\protect\citeauthoryear{{Fialkov}, {Barkana}, {Visbal},
  {Tseliakhovich}  \& {Hirata}}{{Fialkov} et~al.}{2013}]{Fialkov_lyw_2013}
{Fialkov} A.,  {Barkana} R.,  {Visbal} E.,  {Tseliakhovich} D.,   {Hirata}
  C.~M.,  2013, \mn@doi [\mnras] {10.1093/mnras/stt650}, \href
  {https://ui.adsabs.harvard.edu/abs/2013MNRAS.432.2909F} {432, 2909}

\bibitem[\protect\citeauthoryear{{Fialkov}, {Barkana}  \& {Visbal}}{{Fialkov}
  et~al.}{2014}]{Fialkov_2014}
{Fialkov} A.,  {Barkana} R.,   {Visbal} E.,  2014, \mn@doi [Nature]
  {10.1038/nature12999}, \href
  {https://ui.adsabs.harvard.edu/abs/2014Natur.506..197F} {506, 197}

\bibitem[\protect\citeauthoryear{{Fialkov}, {Barkana}  \& {Cohen}}{{Fialkov}
  et~al.}{2018}]{Fialkov2018}
{Fialkov} A.,  {Barkana} R.,   {Cohen} A.,  2018, \mn@doi [prl]
  {10.1103/PhysRevLett.121.011101}, \href
  {https://ui.adsabs.harvard.edu/abs/2018PhRvL.121a1101F} {121, 011101}

\bibitem[\protect\citeauthoryear{{Field}}{{Field}}{1959}]{Field}
{Field} G.~B.,  1959, \mn@doi [ApJ] {10.1086/146653}, 129, 536

\bibitem[\protect\citeauthoryear{{Fixsen} et~al.,}{{Fixsen}
  et~al.}{2011}]{fixsen11}
{Fixsen} D.~J.,  et~al., 2011, \mn@doi [\apj] {10.1088/0004-637X/734/1/5},
  \href {https://ui.adsabs.harvard.edu/abs/2011ApJ...734....5F} {734, 5}

\bibitem[\protect\citeauthoryear{{Foreman-Mackey}, {Hogg}, {Lang}  \&
  {Goodman}}{{Foreman-Mackey} et~al.}{2013}]{Foreman_Mackey_2013}
{Foreman-Mackey} D.,  {Hogg} D.~W.,  {Lang} D.,   {Goodman} J.,  2013, \mn@doi
  [\pasp] {10.1086/670067}, \href
  {https://ui.adsabs.harvard.edu/abs/2013PASP..125..306F} {125, 306}

\bibitem[\protect\citeauthoryear{Fragos, Lehmer, Naoz, Zezas  \&
  Basu-Zych}{Fragos et~al.}{2013}]{Fragos_Xrays_2013}
Fragos T.,  Lehmer B.~D.,  Naoz S.,  Zezas A.,   Basu-Zych A.,  2013, \mn@doi
  [The Astrophysical Journal] {10.1088/2041-8205/776/2/l31}, 776, L31

\bibitem[\protect\citeauthoryear{{Furlanetto}, {Oh}  \& {Briggs}}{{Furlanetto}
  et~al.}{2006}]{Furlanetto_review_2006}
{Furlanetto} S.~R.,  {Oh} S.~P.,   {Briggs} F.~H.,  2006, \mn@doi [\physrep]
  {10.1016/j.physrep.2006.08.002}, \href
  {https://ui.adsabs.harvard.edu/abs/2006PhR...433..181F} {433, 181}

\bibitem[\protect\citeauthoryear{{Garsden} et~al.,}{{Garsden}
  et~al.}{2021}]{LEDA_PS_2021}
{Garsden} H.,  et~al., 2021, \mn@doi [\mnras] {10.1093/mnras/stab1671}, \href
  {https://ui.adsabs.harvard.edu/abs/2021MNRAS.506.5802G} {506, 5802}

\bibitem[\protect\citeauthoryear{Gehlot et~al.,}{Gehlot
  et~al.}{2020}]{AARTFAAC_2020}
Gehlot B.~K.,  et~al., 2020, \mn@doi [\mnras] {10.1093/mnras/staa3093}, 499,
  4158

\bibitem[\protect\citeauthoryear{Greig \& Mesinger}{Greig \&
  Mesinger}{2017}]{greig2017simultaneously}
Greig B.,  Mesinger A.,  2017, Monthly Notices of the Royal Astronomical
  Society, 472, 2651

\bibitem[\protect\citeauthoryear{{Greig}, {Mesinger}  \& {Koopmans}}{{Greig}
  et~al.}{2020}]{Greig_SKA_2020}
{Greig} B.,  {Mesinger} A.,   {Koopmans} L. V.~E.,  2020, \mn@doi [\mnras]
  {10.1093/mnras/stz3138}, \href
  {https://ui.adsabs.harvard.edu/abs/2020MNRAS.491.1398G} {491, 1398}

\bibitem[\protect\citeauthoryear{{Hainline} et~al.,}{{Hainline}
  et~al.}{2023}]{JADES_highz_2023}
{Hainline} K.~N.,  et~al., 2023, \mn@doi [arXiv e-prints]
  {10.48550/arXiv.2306.02468}, \href
  {https://ui.adsabs.harvard.edu/abs/2023arXiv230602468H} {p. arXiv:2306.02468}

\bibitem[\protect\citeauthoryear{Handley}{Handley}{2018}]{fgivenx}
Handley W.,  2018, \mn@doi [The Journal of Open Source Software]
  {10.21105/joss.00849}, 3

\bibitem[\protect\citeauthoryear{Handley, Hobson  \& Lasenby}{Handley
  et~al.}{2015a}]{Handley2015a}
Handley W.~J.,  Hobson M.~P.,   Lasenby A.~N.,  2015a, \mn@doi [MNRAS: Letters]
  {10.1093/mnrasl/slv047}, 450, L61

\bibitem[\protect\citeauthoryear{Handley, Hobson  \& Lasenby}{Handley
  et~al.}{2015b}]{Handley2015b}
Handley W.~J.,  Hobson M.~P.,   Lasenby A.~N.,  2015b, \mn@doi [\mnras]
  {10.1093/mnras/stv1911}, 453, 4385

\bibitem[\protect\citeauthoryear{{Hills}, {Kulkarni}, {Meerburg}  \&
  {Puchwein}}{{Hills} et~al.}{2018}]{Hills2018}
{Hills} R.,  {Kulkarni} G.,  {Meerburg} P.~D.,   {Puchwein} E.,  2018, \mn@doi
  [Nature] {10.1038/s41586-018-0796-5}, \href
  {https://ui.adsabs.harvard.edu/abs/2018Natur.564E..32H} {564, E32}

\bibitem[\protect\citeauthoryear{{Jacobs} et~al.,}{{Jacobs}
  et~al.}{2015}]{Jacobs_Paper_limits_2015}
{Jacobs} D.~C.,  et~al., 2015, \mn@doi [\apj] {10.1088/0004-637X/801/1/51},
  \href {https://ui.adsabs.harvard.edu/abs/2015ApJ...801...51J} {801, 51}

\bibitem[\protect\citeauthoryear{Jana, Nath  \& Biermann}{Jana
  et~al.}{2019}]{Jana2018}
Jana R.,  Nath B.~B.,   Biermann P.~L.,  2019, \mn@doi [\mnras]
  {10.1093/mnras/sty3426}, 483, 5329

\bibitem[\protect\citeauthoryear{{Klessen}}{{Klessen}}{2019}]{Klessen_pop3_2019}
{Klessen} R.,  2019, in {Latif} M.,  {Schleicher} D.,  eds, , Formation of the
  First Black Holes.
pp 67--97, \mn@doi{10.1142/9789813227958_0004}

\bibitem[\protect\citeauthoryear{{Kolopanis}, {Pober}, {Jacobs}  \&
  {McGraw}}{{Kolopanis} et~al.}{2022}]{Kolopanis_MWA_limits_2022}
{Kolopanis} M.,  {Pober} J.,  {Jacobs} D.~C.,   {McGraw} S.,  2022, arXiv
  e-prints, \href {https://ui.adsabs.harvard.edu/abs/2022arXiv221010885K} {p.
  arXiv:2210.10885}

\bibitem[\protect\citeauthoryear{Koopmans}{Koopmans}{2017}]{LOFAR_EoR_2018}
Koopmans L. V.~E.,  2017, \mn@doi [Proceedings of the International
  Astronomical Union] {10.1017/S1743921318000583}, 12, 71–76

\bibitem[\protect\citeauthoryear{{Koopmans} et~al.,}{{Koopmans}
  et~al.}{2015}]{Koopmans_SKA_2015}
{Koopmans} L.,  et~al., 2015, in Advancing Astrophysics with the Square
  Kilometre Array (AASKA14). p.~1 (\mn@eprint {arXiv} {1505.07568}),
  \mn@doi{10.22323/1.215.0001}

\bibitem[\protect\citeauthoryear{{Kovetz}, {Poulin}, {Gluscevic}, {Boddy},
  {Barkana}  \& {Kamionkowski}}{{Kovetz} et~al.}{2018}]{Kovetz2018}
{Kovetz} E.~D.,  {Poulin} V.,  {Gluscevic} V.,  {Boddy} K.~K.,  {Barkana} R.,
  {Kamionkowski} M.,  2018, \mn@doi [Phys. Rev.~D]
  {10.1103/PhysRevD.98.103529}, 98, 103529

\bibitem[\protect\citeauthoryear{Kullback \& Leibler}{Kullback \&
  Leibler}{1951}]{kullback_information_1951}
Kullback S.,  Leibler R.~A.,  1951, \mn@doi [The Annals of Mathematical
  Statistics] {10.1214/aoms/1177729694}, 22, 79

\bibitem[\protect\citeauthoryear{{Lehmer} et~al.,}{{Lehmer}
  et~al.}{2012}]{Lehmer_Chandra_2012}
{Lehmer} B.~D.,  et~al., 2012, \mn@doi [\apj] {10.1088/0004-637X/752/1/46},
  \href {https://ui.adsabs.harvard.edu/abs/2012ApJ...752...46L} {752, 46}

\bibitem[\protect\citeauthoryear{{Li}, {Houston}, {Li}, {Yang}  \&
  {Zhang}}{{Li} et~al.}{2021}]{axion_dm_2021}
{Li} C.,  {Houston} N.,  {Li} T.,  {Yang} Q.,   {Zhang} X.,  2021, \mn@doi
  [International Journal of Modern Physics D] {10.1142/S0218271821500413},
  \href {https://ui.adsabs.harvard.edu/abs/2021IJMPD..3050041L} {30, 2150041}

\bibitem[\protect\citeauthoryear{Liu, Outmezguine, Redigolo  \& Volansky}{Liu
  et~al.}{2019}]{Liu2019}
Liu H.,  Outmezguine N.~J.,  Redigolo D.,   Volansky T.,  2019, \mn@doi [Phys.
  Rev. D] {10.1103/PhysRevD.100.123011}, 100, 123011

\bibitem[\protect\citeauthoryear{{Madau}, {Meiksin}  \& {Rees}}{{Madau}
  et~al.}{1997}]{Madau}
{Madau} P.,  {Meiksin} A.,   {Rees} M.~J.,  1997, \mn@doi [\apj]
  {10.1086/303549}, \href
  {https://ui.adsabs.harvard.edu/abs/1997ApJ...475..429M} {475, 429}

\bibitem[\protect\citeauthoryear{{Mellema} et~al.,}{{Mellema}
  et~al.}{2013}]{Mellema_SKA_2013}
{Mellema} G.,  et~al., 2013, \mn@doi [Experimental Astronomy]
  {10.1007/s10686-013-9334-5}, \href
  {https://ui.adsabs.harvard.edu/abs/2013ExA....36..235M} {36, 235}

\bibitem[\protect\citeauthoryear{{Mertens} et~al.,}{{Mertens}
  et~al.}{2020}]{Mertens_2020}
{Mertens} F.~G.,  et~al., 2020, \mn@doi [\mnras] {10.1093/mnras/staa327}, \href
  {https://ui.adsabs.harvard.edu/abs/2020MNRAS.493.1662M} {493, 1662}

\bibitem[\protect\citeauthoryear{{Mertens}, {Semelin}  \& {Koopmans}}{{Mertens}
  et~al.}{2021}]{Mertens_NenuFAR_2021}
{Mertens} F.~G.,  {Semelin} B.,   {Koopmans} L.~V.~E.,  2021, in {Siebert} A.,
  et~al., eds, SF2A-2021: Proceedings of the Annual meeting of the French
  Society of Astronomy and Astrophysics. pp 211--214 (\mn@eprint {arXiv}
  {2109.10055})

\bibitem[\protect\citeauthoryear{{Mesinger}}{{Mesinger}}{2019}]{Mesinger_review_2019}
{Mesinger} A.,  2019, {The Cosmic 21-cm Revolution; Charting the first billion
  years of our universe}, \mn@doi{10.1088/2514-3433/ab4a73.
}

\bibitem[\protect\citeauthoryear{{Mesinger}, {Furlanetto}  \& {Cen}}{{Mesinger}
  et~al.}{2011}]{Mesinger_21cmFAST_2011}
{Mesinger} A.,  {Furlanetto} S.,   {Cen} R.,  2011, \mn@doi [\mnras]
  {10.1111/j.1365-2966.2010.17731.x}, \href
  {https://ui.adsabs.harvard.edu/abs/2011MNRAS.411..955M} {411, 955}

\bibitem[\protect\citeauthoryear{{Mirocha} \& {Furlanetto}}{{Mirocha} \&
  {Furlanetto}}{2019}]{Mirocha2019}
{Mirocha} J.,  {Furlanetto} S.~R.,  2019, \mn@doi [\mnras]
  {10.1093/mnras/sty3260}, 483, 1980

\bibitem[\protect\citeauthoryear{{Mirocha}, {Furlanetto}  \& {Sun}}{{Mirocha}
  et~al.}{2017}]{Mirocha_global_2017}
{Mirocha} J.,  {Furlanetto} S.~R.,   {Sun} G.,  2017, \mn@doi [\mnras]
  {10.1093/mnras/stw2412}, \href
  {https://ui.adsabs.harvard.edu/abs/2017MNRAS.464.1365M} {464, 1365}

\bibitem[\protect\citeauthoryear{{Mittal} \& {Kulkarni}}{{Mittal} \&
  {Kulkarni}}{2022a}]{Mittal2022}
{Mittal} S.,  {Kulkarni} G.,  2022a, \mn@doi [\mnras] {10.1093/mnras/stac1961},
  \href {https://ui.adsabs.harvard.edu/abs/2022MNRAS.tmp.1856M} {}

\bibitem[\protect\citeauthoryear{{Mittal} \& {Kulkarni}}{{Mittal} \&
  {Kulkarni}}{2022b}]{Mittal_ERB_2022}
{Mittal} S.,  {Kulkarni} G.,  2022b, \mn@doi [\mnras] {10.1093/mnras/stac005},
  \href {https://ui.adsabs.harvard.edu/abs/2022MNRAS.510.4992M} {510, 4992}

\bibitem[\protect\citeauthoryear{{Mondal} et~al.,}{{Mondal}
  et~al.}{2020}]{Mondal_LOFAR_2022}
{Mondal} R.,  et~al., 2020, \mn@doi [\mnras] {10.1093/mnras/staa2422}, \href
  {https://ui.adsabs.harvard.edu/abs/2020MNRAS.498.4178M} {498, 4178}

\bibitem[\protect\citeauthoryear{{Monsalve}, {Rogers}, {Bowman}  \&
  {Mozdzen}}{{Monsalve} et~al.}{2017}]{Monsalve_EDGES_HB_1_2017}
{Monsalve} R.~A.,  {Rogers} A. E.~E.,  {Bowman} J.~D.,   {Mozdzen} T.~J.,
  2017, \mn@doi [\apj] {10.3847/1538-4357/aa88d1}, \href
  {https://ui.adsabs.harvard.edu/abs/2017ApJ...847...64M} {847, 64}

\bibitem[\protect\citeauthoryear{{Monsalve}, {Greig}, {Bowman}, {Mesinger},
  {Rogers}, {Mozdzen}, {Kern}  \& {Mahesh}}{{Monsalve}
  et~al.}{2018}]{Monsalve_EDGES_HB_2_2018}
{Monsalve} R.~A.,  {Greig} B.,  {Bowman} J.~D.,  {Mesinger} A.,  {Rogers} A.
  E.~E.,  {Mozdzen} T.~J.,  {Kern} N.~S.,   {Mahesh} N.,  2018, \mn@doi [\apj]
  {10.3847/1538-4357/aace54}, \href
  {https://ui.adsabs.harvard.edu/abs/2018ApJ...863...11M} {863, 11}

\bibitem[\protect\citeauthoryear{{Monsalve}, {Fialkov}, {Bowman}, {Rogers},
  {Mozdzen}, {Cohen}, {Barkana}  \& {Mahesh}}{{Monsalve}
  et~al.}{2019}]{Monsalve_EDGES_HB_3_2019}
{Monsalve} R.~A.,  {Fialkov} A.,  {Bowman} J.~D.,  {Rogers} A. E.~E.,
  {Mozdzen} T.~J.,  {Cohen} A.,  {Barkana} R.,   {Mahesh} N.,  2019, \mn@doi
  [\apj] {10.3847/1538-4357/ab07be}, \href
  {https://ui.adsabs.harvard.edu/abs/2019ApJ...875...67M} {875, 67}

\bibitem[\protect\citeauthoryear{{Monsalve} et~al.,}{{Monsalve}
  et~al.}{2023}]{MIST}
{Monsalve} R.~A.,  et~al., 2023, \mn@doi [arXiv e-prints]
  {10.48550/arXiv.2309.02996}, \href
  {https://ui.adsabs.harvard.edu/abs/2023arXiv230902996M} {p. arXiv:2309.02996}

\bibitem[\protect\citeauthoryear{{Mu{\~n}oz} \& {Loeb}}{{Mu{\~n}oz} \&
  {Loeb}}{2018}]{Munoz2018}
{Mu{\~n}oz} J.~B.,  {Loeb} A.,  2018, \mn@doi [Nature]
  {10.1038/s41586-018-0151-x}, 557, 684

\bibitem[\protect\citeauthoryear{{Mu{\~n}oz}, {Qin}, {Mesinger}, {Murray},
  {Greig}  \& {Mason}}{{Mu{\~n}oz} et~al.}{2022}]{Munoz_first_gals_2022}
{Mu{\~n}oz} J.~B.,  {Qin} Y.,  {Mesinger} A.,  {Murray} S.~G.,  {Greig} B.,
  {Mason} C.,  2022, \mn@doi [\mnras] {10.1093/mnras/stac185}, \href
  {https://ui.adsabs.harvard.edu/abs/2022MNRAS.511.3657M} {511, 3657}

\bibitem[\protect\citeauthoryear{{Mushotzky}}{{Mushotzky}}{2018}]{axis}
{Mushotzky} R.,  2018, in {den Herder} J.-W.~A.,  {Nikzad} S.,   {Nakazawa} K.,
   eds,  Society of Photo-Optical Instrumentation Engineers (SPIE) Conference
  Series Vol. 10699, Space Telescopes and Instrumentation 2018: Ultraviolet to
  Gamma Ray. p. 1069929 (\mn@eprint {arXiv} {1807.02122}),
  \mn@doi{10.1117/12.2310003}

\bibitem[\protect\citeauthoryear{{Naidu} et~al.,}{{Naidu}
  et~al.}{2022}]{Naidu_2022}
{Naidu} R.~P.,  et~al., 2022, \mn@doi [arXiv e-prints]
  {10.48550/arXiv.2208.02794}, \href
  {https://ui.adsabs.harvard.edu/abs/2022arXiv220802794N} {p. arXiv:2208.02794}

\bibitem[\protect\citeauthoryear{{Park}, {Gillet}, {Mesinger}  \&
  {Greig}}{{Park} et~al.}{2020}]{Park_JWST_2020}
{Park} J.,  {Gillet} N.,  {Mesinger} A.,   {Greig} B.,  2020, \mn@doi [\mnras]
  {10.1093/mnras/stz3278}, \href
  {https://ui.adsabs.harvard.edu/abs/2020MNRAS.491.3891P} {491, 3891}

\bibitem[\protect\citeauthoryear{{Patil} et~al.,}{{Patil}
  et~al.}{2017}]{Patil_2017}
{Patil} A.~H.,  et~al., 2017, \mn@doi [\apj] {10.3847/1538-4357/aa63e7}, \href
  {https://ui.adsabs.harvard.edu/abs/2017ApJ...838...65P} {838, 65}

\bibitem[\protect\citeauthoryear{{Philip} et~al.,}{{Philip}
  et~al.}{2019}]{Philip_prizm_2019}
{Philip} L.,  et~al., 2019, \mn@doi [Journal of Astronomical Instrumentation]
  {10.1142/S2251171719500041}, \href
  {https://ui.adsabs.harvard.edu/abs/2019JAI.....850004P} {8, 1950004}

\bibitem[\protect\citeauthoryear{{Planck Collaboration} et~al.,}{{Planck
  Collaboration} et~al.}{2020}]{Planck2018}
{Planck Collaboration} et~al., 2020, \mn@doi [A\&A]
  {10.1051/0004-6361/201833910}, \href
  {https://ui.adsabs.harvard.edu/abs/2020A&A...641A...6P} {641, A6}

\bibitem[\protect\citeauthoryear{{Reis}, {Fialkov}  \& {Barkana}}{{Reis}
  et~al.}{2020}]{Reis2020}
{Reis} I.,  {Fialkov} A.,   {Barkana} R.,  2020, \mn@doi [\mnras]
  {10.1093/mnras/staa3091}, \href
  {https://ui.adsabs.harvard.edu/abs/2020MNRAS.499.5993R} {499, 5993}

\bibitem[\protect\citeauthoryear{{Reis}, {Fialkov}  \& {Barkana}}{{Reis}
  et~al.}{2021}]{Reis_lya_2021}
{Reis} I.,  {Fialkov} A.,   {Barkana} R.,  2021, \mn@doi [\mnras]
  {10.1093/mnras/stab2089}, \href
  {https://ui.adsabs.harvard.edu/abs/2021MNRAS.506.5479R} {506, 5479}

\bibitem[\protect\citeauthoryear{{Robertson}}{{Robertson}}{2022}]{Robertson_JWST_2022}
{Robertson} B.~E.,  2022, \mn@doi [\araa]
  {10.1146/annurev-astro-120221-044656}, \href
  {https://ui.adsabs.harvard.edu/abs/2022ARA&A..60..121R} {60, 121}

\bibitem[\protect\citeauthoryear{{Robertson} et~al.,}{{Robertson}
  et~al.}{2023}]{Robertson_2023}
{Robertson} B.~E.,  et~al., 2023, \mn@doi [Nature Astronomy]
  {10.1038/s41550-023-01921-1}, \href
  {https://ui.adsabs.harvard.edu/abs/2023NatAs...7..611R} {7, 611}

\bibitem[\protect\citeauthoryear{{Sims} \& {Pober}}{{Sims} \&
  {Pober}}{2020}]{Sims2020}
{Sims} P.~H.,  {Pober} J.~C.,  2020, \mn@doi [\mnras] {10.1093/mnras/stz3388},
  \href {https://ui.adsabs.harvard.edu/abs/2020MNRAS.492...22S} {492, 22}

\bibitem[\protect\citeauthoryear{{Singh} \& {Subrahmanyan}}{{Singh} \&
  {Subrahmanyan}}{2019}]{Singh2019}
{Singh} S.,  {Subrahmanyan} R.,  2019, \mn@doi [ApJ]
  {10.3847/1538-4357/ab2879}, \href
  {https://ui.adsabs.harvard.edu/abs/2019ApJ...880...26S} {880, 26}

\bibitem[\protect\citeauthoryear{{Singh} et~al.,}{{Singh}
  et~al.}{2017}]{SARAS2_2017}
{Singh} S.,  et~al., 2017, \mn@doi [\apjl] {10.3847/2041-8213/aa831b}, \href
  {https://ui.adsabs.harvard.edu/abs/2017ApJ...845L..12S} {845, L12}

\bibitem[\protect\citeauthoryear{{Singh} et~al.,}{{Singh}
  et~al.}{2018}]{SARAS2_2018}
{Singh} S.,  et~al., 2018, \mn@doi [ApJ] {10.3847/1538-4357/aabae1}, \href
  {https://ui.adsabs.harvard.edu/abs/2018ApJ...858...54S} {858, 54}

\bibitem[\protect\citeauthoryear{{Singh} et~al.,}{{Singh}
  et~al.}{2022}]{SARAS3}
{Singh} S.,  et~al., 2022, \mn@doi [Nature Astronomy]
  {10.1038/s41550-022-01610-5}, \href
  {https://ui.adsabs.harvard.edu/abs/2022NatAs...6..607S} {6, 607}

\bibitem[\protect\citeauthoryear{Skilling}{Skilling}{2004}]{skilling_nested_2004}
Skilling J.,  2004, \mn@doi [AIP Conference Proceedings] {10.1063/1.1835238},
  735, 395

\bibitem[\protect\citeauthoryear{{Slatyer} \& {Wu}}{{Slatyer} \&
  {Wu}}{2018}]{Slatyer2018}
{Slatyer} T.~R.,  {Wu} C.,  2018, \mn@doi [Phys. Rev.~D]
  {10.1103/PhysRevD.98.023013}, 98, 023013

\bibitem[\protect\citeauthoryear{{The Athena Collaboration}}{{The Athena
  Collaboration}}{2013}]{Athena}
{The Athena Collaboration} 2013, arXiv e-prints, \href
  {https://ui.adsabs.harvard.edu/abs/2013arXiv1306.2307N} {p. arXiv:1306.2307}

\bibitem[\protect\citeauthoryear{{The HERA Collaboration} et~al.,}{{The HERA
  Collaboration} et~al.}{2022}]{HERA_2022c}
{The HERA Collaboration} et~al., 2022, arXiv e-prints, \href
  {https://ui.adsabs.harvard.edu/abs/2022arXiv221004912T} {p. arXiv:2210.04912}

\bibitem[\protect\citeauthoryear{{The Lynx Team}}{{The Lynx Team}}{2018}]{Lynx}
{The Lynx Team} 2018, arXiv e-prints, \href
  {https://ui.adsabs.harvard.edu/abs/2018arXiv180909642T} {p. arXiv:1809.09642}

\bibitem[\protect\citeauthoryear{{Trott} et~al.,}{{Trott}
  et~al.}{2020}]{Trott_2020}
{Trott} C.~M.,  et~al., 2020, \mn@doi [\mnras] {10.1093/mnras/staa414}, \href
  {https://ui.adsabs.harvard.edu/abs/2020MNRAS.493.4711T} {493, 4711}

\bibitem[\protect\citeauthoryear{{Venumadhav}, {Dai}, {Kaurov}  \&
  {Zaldarriaga}}{{Venumadhav} et~al.}{2018}]{Venumadhav2018}
{Venumadhav} T.,  {Dai} L.,  {Kaurov} A.,   {Zaldarriaga} M.,  2018, \mn@doi
  [Phys.\ Rev.~D] {10.1103/PhysRevD.98.103513}, \href
  {https://ui.adsabs.harvard.edu/abs/2018PhRvD..98j3513V} {98, 103513}

\bibitem[\protect\citeauthoryear{{Visbal}, {Barkana}, {Fialkov},
  {Tseliakhovich}  \& {Hirata}}{{Visbal} et~al.}{2012}]{Visbal_2012}
{Visbal} E.,  {Barkana} R.,  {Fialkov} A.,  {Tseliakhovich} D.,   {Hirata}
  C.~M.,  2012, \mn@doi [Nature] {10.1038/nature11177}, \href
  {https://ui.adsabs.harvard.edu/abs/2012Natur.487...70V} {487, 70}

\bibitem[\protect\citeauthoryear{{Windhorst}, {Cohen}, {Jansen}, {Conselice}
  \& {Yan}}{{Windhorst} et~al.}{2006}]{Windhorst_JWST_2006}
{Windhorst} R.~A.,  {Cohen} S.~H.,  {Jansen} R.~A.,  {Conselice} C.,   {Yan}
  H.,  2006, \mn@doi [Nature] {10.1016/j.newar.2005.11.018}, \href
  {https://ui.adsabs.harvard.edu/abs/2006NewAR..50..113W} {50, 113}

\bibitem[\protect\citeauthoryear{{Wouthuysen}}{{Wouthuysen}}{1952}]{Wouthuysen}
{Wouthuysen} S.~A.,  1952, \mn@doi [\aj] {10.1086/106661}, 57, 31

\bibitem[\protect\citeauthoryear{{Wu}, {Xu}  \& {Zheng}}{{Wu}
  et~al.}{2022}]{Wimpless_DM_2022}
{Wu} S.,  {Xu} S.,   {Zheng} S.,  2022, \mn@doi [arXiv e-prints]
  {10.48550/arXiv.2205.14876}, \href
  {https://ui.adsabs.harvard.edu/abs/2022arXiv220514876W} {p. arXiv:2205.14876}

\bibitem[\protect\citeauthoryear{Zarka, Coffre, Denis, Dumez-Viou, Girard,
  Grießmeier, Loh  \& Tagger}{Zarka et~al.}{2018}]{Zarka_nenuFar_2018}
Zarka P.,  Coffre A.,  Denis L.,  Dumez-Viou C.,  Girard J.,  Grießmeier
  J.-M.,  Loh A.,   Tagger M.,  2018, in 2018 2nd URSI Atlantic Radio Science
  Meeting (AT-RASC). pp~1--1, \mn@doi{10.23919/URSI-AT-RASC.2018.8471648}

\bibitem[\protect\citeauthoryear{{de Lera Acedo} et~al.,}{{de Lera Acedo}
  et~al.}{2022}]{REACH}
{de Lera Acedo} E.,  et~al., 2022, \mn@doi [Nature Astronomy]
  {10.1038/s41550-022-01709-9}, \href
  {https://ui.adsabs.harvard.edu/abs/2022NatAs...6..984D} {6, 984}

\makeatother
\end{thebibliography}
\bibliographystyle{mnras}

%%%%%%%%%%%%%%%%%%%%%%%%%%%%%%%%%%%%%%%%%%%%%%%%%%

%%%%%%%%%%%%%%%%% APPENDICES %%%%%%%%%%%%%%%%%%%%%

\appendix

\section{Individual Constraints from HERA and SARAS3 on Key Parameters}
\label{app:key-params}

\begin{figure*}
    \centering
    \includegraphics{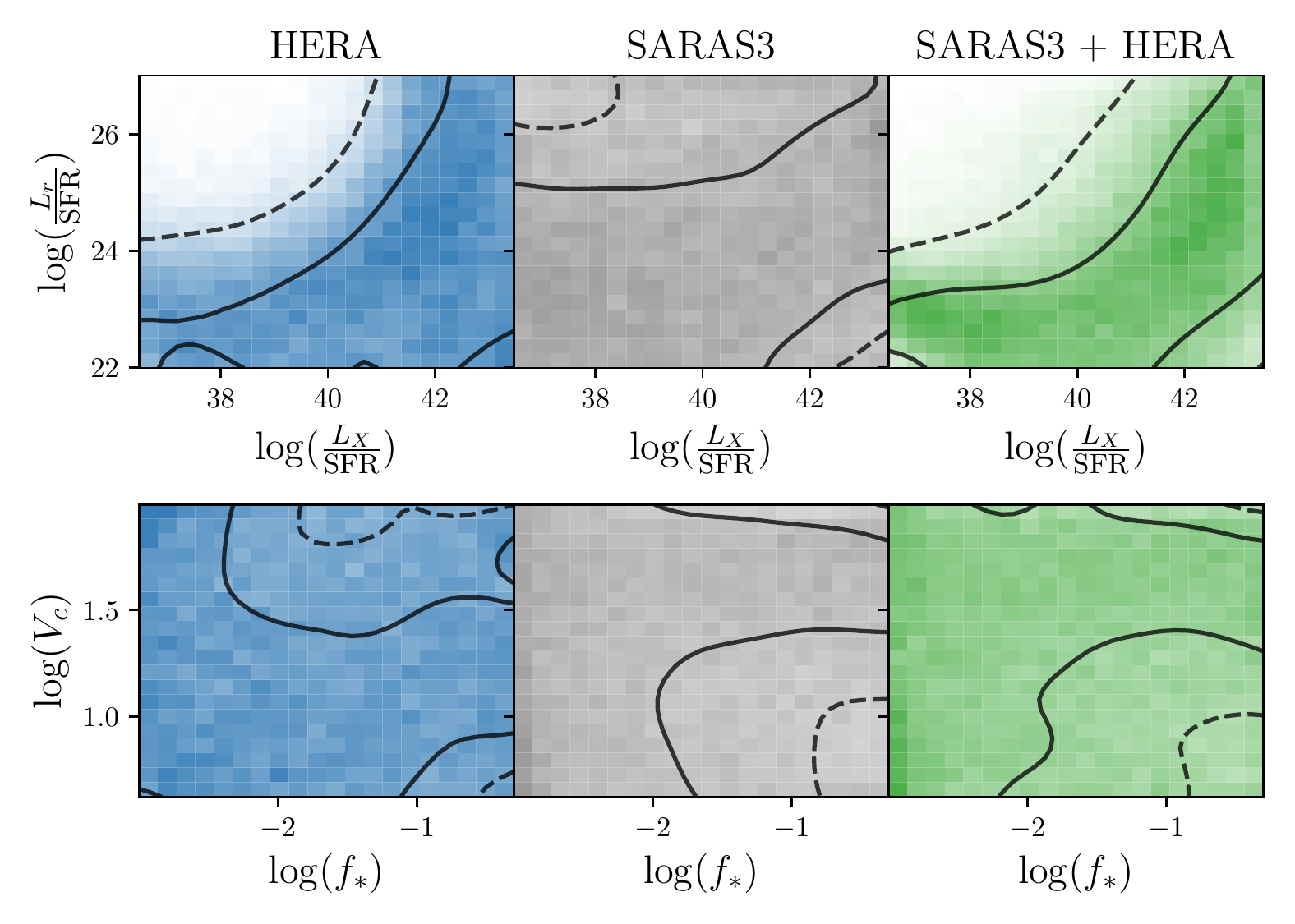}
    \caption{The figure shows the individual constraints in the $\L_X$ vs $L_r$ and $V_c$ vs $f_*$ parameter spaces, marginalised over the other parameters in the model, from SARAS3 and HERA along with the joint analysis. HERA dominates the constraints on the radio and X-ray luminosities, but SARAS3 disfavours an additional part of the parameter space corresponding to higher $L_X$ and high $L_r$. In addition, we see that SARAS3 dominates the constraints on the properties that govern star formation, $V_c$ and $f_*$, in the joint analysis because the higher redshifts covered by the SARAS3 band are more sensitive to these processes.}
    \label{fig:overlay_hist}
\end{figure*}

In \cref{fig:overlay_hist}, we show the individual constraints from HERA and SARAS3 along with the joint constraints on the two constrained parts of the parameter space $L_X$ vs $L_r$ and $V_c$ vs $f_*$. All other parameters have been marginalised out in these plots.

From the figure, we see that HERA dominates the constraints on the strength of the radio and X-ray backgrounds. However, SARAS3 provides disfavours at 68\% confidence higher $L_X$ in combination with high $L_r$ than HERA leading to a tighter constraint on the parameter space in the joint analysis.

For the properties governing star formation, namely $f_*$ and $V_c$, we see that SARAS3 dominates the joint analysis. The SARAS3 observations are at higher redshifts than HERA and cover a period of Cosmic history where the 21-cm signal is more sensitive to the processes of star formation. The apparent constraint from HERA on the star formation properties is washed out by SARAS3 since SARAS3 is more constraining.

The figure shows how the two probes are complimentary, providing additional information because they cover different redshifts dominated by different physical processes.

% Don't change these lines
\bsp	% typesetting comment
\label{lastpage}
\end{document}